\pdfoutput=1
\documentclass[11pt]{article}

\usepackage{color}
\usepackage{graphicx}
\usepackage{amsmath}
\usepackage{amssymb}
\usepackage{dcolumn}
\usepackage{bm}
\usepackage{slashed}
\usepackage{authblk}
\usepackage{textcomp}
\usepackage{cite}

\newcommand{\be}{\begin{equation}}
\newcommand{\ee}{\end{equation}}
\newcommand{\ba}{\begin{eqnarray}}
\newcommand{\ea}{\end{eqnarray}}
\newcommand{\no}{\nonumber \\}
\newcommand{\gsim}{\mathrel{\hbox{\rlap{\lower.55ex \hbox {$\sim$}}
                   \kern-.3em \raise.4ex \hbox{$>$}}}}
\newcommand{\lsim}{\mathrel{\hbox{\rlap{\lower.55ex \hbox {$\sim$}}
                   \kern-.3em \raise.4ex \hbox{$<$}}}}

\def\roughly#1{\mathrel{\raise.3ex\hbox{$#1$\kern-.75em%
\lower1ex\hbox{$\sim$}}}}
\def\lsim{\roughly<}
\def\gsim{\roughly>}

\def\({\left(}
\def\){\right)}
\def\[{\left[}
\def\]{\right]}
\def\<{\langle}
\def\>{\rangle}

\def\tr{\text{tr}}

\def\tm{{\tilde \mu}}
\def\bmu{\bar{\mu}}

\def\h{{\eta}}

\def\l{{\lambda}}
\def\L{{\Lambda}}
\def\d{{\delta}}
\def\D{{\Delta}}

\def\e{{\epsilon}}

\def\a{{\alpha}}
\def\b{{\beta}}
\def\c{{\chi}}

\def\g{{\gamma}}
\def\G{{\Gamma}}
\def\p{{\pi}}
\def\P{{\Pi}}
\def\m{{\mu}}
\def\n{{\nu}}
\def\r{{\rho}}
\def\s{{\sigma}}
\def\S{{\Sigma}}
\def\t{{\tau}}
\def\th{{\theta}}
\def\ph{{\phi}}

\def\P{{\Pi}}

\def\quv{q_\text{UV}}
\def\qir{q_\text{IR}}

\newcommand{\pd}{{\partial}}

\newcommand{\sgn}{\text{sgn}}

\date{\today}

\begin{document}

\title{\bf Polarization Dynamics in Paramagnet of Charged Quark-Gluon Plasma}

\author[1]{{Lihua Dong}
	\thanks{donglh6@mail2.sysu.edu.cn}}
\author[1]{{Shu Lin}
	\thanks{linshu8@mail.sysu.edu.cn}}
\affil[1]{School of Physics and Astronomy, Sun Yat-Sen University, Zhuhai 519082, China}

\maketitle

\begin{abstract}
It is commonly understood that the strong magnetic field produced in heavy ion collisions is short-lived. The electric conductivity of the quark-gluon plasma is unable to significantly extend the life time of magnetic field. We propose an alternative scenario to achieve this: with finite baryon density and spin polarization by the initial magnetic field, the quark-gluon plasma behaves as a paramagnet, which may continue to polarize quark after fading of initial magnetic field. We confirm this picture by calculations in both quantum electrodynamics and quantum chromodynamics. In the former case, we find a splitting in the damping rates of probe fermion with opposite spin component along the magnetic field. In the latter case, we find a similar splitting in damping rate of probe quark in quark-gluon plasma in both high density and low density limits. The splitting provides a way of polarizing strange quarks by the quark-gluon plasma paramagnet consisting of light quarks, which effectively extends the lifetime of magnetic field in heavy ion collisions.
\end{abstract}


\newpage

\section{Introduction}

The observations of spin polarization in $\L$-hyperon in heavy ion collision experiments have revealed quark-gluon plasma (QGP) as spin polarized matter \cite{STAR:2017ckg}. The polarization is attributed to vorticity of QGP coming from initial orbital angular momentum in off-central collisions \cite{Liang:2004ph}. Theories based on spin-vorticity coupling have been developed in the past few years \cite{Becattini:2007nd,Becattini:2013fla,Fang:2016vpj,Becattini:2018duy,Gao:2023wwo}, giving satisfactory explanation of global spin polarization \cite{Karpenko:2016jyx,Sun:2017xhx,Li:2017slc,Wei:2018zfb,Xie:2017upb}. However, the spin-vorticity coupling alone predicts an equal polarization for both $\L$ and anti-$\L$, while experiments have found splitting of polarizations for $\L$ and anti-$\L$, with the splitting more prominent at low energy collisions. Different mechanisms have been proposed to understand the splitting including spin-magnetic coupling \cite{Becattini:2016gvu,Guo:2019joy,Xu:2022hql}, mean-field effect \cite{Csernai:2018yok}, direct flow effect \cite{Jiang:2023fad}, helicity vortical effect \cite{Ambrus:2020oiw} etc.

While the mechanism of spin-magnetic coupling gives the correct sign of polarization splitting, it is generally expected that it cannot provide sufficient magnitude because the lifetime of the magnetic field is short so that the remaining magnetic field at freezeout may be too weak. Indeed, recent studies suggest magnetic field alone cannot explain the splitting at low energy \cite{Guo:2019joy,Xu:2022hql}. The evolution of magnetic field has been studied using evolution of in-medium electromagnetic field \cite{McLerran:2013hla,Li:2016tel}. In order to have long life time for magnetic field, one needs to have large electric conductivity for QGP medium, which is not favored by lattice studies \cite{Amato:2013naa,Aarts:2014nba,Brandt:2015aqk,Ding:2016hua}. Anisotropic conductivity in magnetized QGP has been considered in different approaches including lattice \cite{Astrakhantsev:2019zkr}, holography \cite{Li:2018ufq,Fukushima:2021got} and kinetic theories \cite{Hattori:2016cnt,Hattori:2016lqx,Fukushima:2019ugr,Fukushima:2017lvb,Lin:2019fqo,Ghosh:2019ubc,Yan:2021zjc,Peng:2023rjj}. However, the situation does not improve significantly at phenomenologically relevant strength of magnetic field. Other methods of constraining the strength of magnetic field experimentally have been discussed in \cite{Xu:2020sui}.

Most previous studies have treated QGP as spinless fluid, which does not develop magnetization under external magnetic field. Indeed this is true for charge neutral QGP, in which the spin polarization due to spin-magnetic coupling cancel among positive and negative charge carriers. However, the cancellation is incomplete in charged QGP, leading to nonvanishing magnetization. This is most clearly seen in strong magnetic field limit, where the fermionic degrees of freedom are dominated by lowest Landau levels (LLL), see \cite{Hattori:2023egw} for a recent review. The magnetization by the LLL leads to net spin polarization in charged QGP \footnote{QGP produced at low energy collisions has net baryon charge. It is also electrically charged for two-flavor QGP.}. In particular, positively charged QGP relevant for heavy ion phenomenology corresponds to a paramagnet.

Recently the magnetic susceptibility of charge neutral quantum chromodynamics (QCD) matter has been studied on the lattice in the weak magnetic field limit \cite{Bali:2014kia,Bali:2020bcn,Levkova:2013qda}, see also \cite{Buividovich:2021fsa} for a study of charged QCD matter. The  matter is found to be paramagnet in the high temperature phase and diamagnet in low temperature phase. 
Here we consider instead charged QGP in strong magnetic field limit. With net charge, we can loosely regard magnetization as spin polarization like in ordinary magnetic materials. The strong magnetic field limit is merely a technical simplification. With the apparent equivalence of magnetization and polarization in mind, we shall refer to charged QGP as a paramagnet\footnote{In fact, the magnetic susceptibility vanishes because the thermodynamic potential is linear in the magnetic field in the LLL approximation. The vanishing of susceptibility is due to cancellation between spin and orbital angular momentum contributions.} and explore its role in dynamics of spin polarization.

We will propose the following picture for magnetic field induced polarization dynamics in heavy ion collisions: while the magnetic field due to spectators in heavy ion collisions decays quickly, the strong magnetic field can convert the charged QGP consisting of light flavors into a paramagnet. The QGP paramagnet continues to polarize the strange quarks produced at later stage in QGP evolution. The polarization is realized as a splitting of damping rates for strange quark with opposite spin component along the magnetic field, which dynamically favors strange quarks with negative spin component.

The paper is organized as follows: in Sec.~\ref{gbpm}, we review photon self-energy in charged fluid consisting of LLL states, and calculate the resummed photon propagator. We shall find an anti-symmetric component unique to charged fluid, which is essential for polarization dynamics; in Sec.~\ref{pfpm}, we consider a probe fermion in the paramagnet and find a splitting in the damping rates of the probe fermion with opposite spin component along the magnetic field. It provides a mechanism for polarizing the probe fermion; in Sec.~\ref{sec-QGP}, we extend the analysis to probe quark in charged QGP. This case is complicated by self-interaction of gluons, which gives rise to completely different dispersion of gluons. Nevertheless, we find the similar mechanism exists for probe quark. We also discuss implications for heavy ion phenomenology; Sec.~\ref{co} is devoted to conclusion and discussion of future directions.

We define $\e^{0123}=+ 1$, $P^{\m}=(p_{0},\bold{p})$, $\s^{\m}=(1,\boldsymbol{\s})$ and $\bar{\s}^{\m}=(1,-\boldsymbol{\s})$.

\section{Photon in paramagnet}\label{gbpm}

In this section, we study the dynamics of photon in charged magnetized plasma. The case for charge neutral plasma has been studied extensively in literature, see \cite{Hattori:2023egw,Hattori:2022uzp,Hattori:2022wao} and references therein. We shall focus on the difference in charged magnetized plasma. On general ground, charged magnetized plasma consisting of spin one half matter is also spin polarized with nonvanishing magnetization. It is known that medium with magnetization is gyrotropic \cite{LANDAU1984331}, which is characterized by polarization tensor with purely imaginary off-diagonal components. It leads to splitting of right-handed and left-handed electromagnetic waves. We shall see this is also true with the paramagnet. We will first present photon self-energy in charged magnetized plasma, which is then used to determined the dispersion of electromagnetic waves. We will also calculate the resummed photon propagator to be used in Sec.~\ref{pfpm}.

\subsection{Photon self-energy in charged magnetized plasma}

We will use the real time formalism of finite temperature field theory in $ra$-basis \cite{Chou:1984es}. The fields in $ra$-basis are related to the counterpart on Schwinger-Keldysh contour by
\begin{align}\label{ra_12}
A_r=\frac{1}{2}\(A_1+A_2\),\quad A_a=A_1-A_2.
\end{align}
The correlators in the $ra$-basis are defined as
\begin{align}\label{D_def}
&D_{ra}^{\m\n}(x)=\<A^\m_r(x)\,A^\n_a(0)\>,\nonumber\\
&D_{ar}^{\m\n}(x)=\<A^\m_a(x)\,A^\n_r(0)\>,\nonumber\\
&D_{rr}^{\m\n}(x)=\<A^\m_r(x)\,A^\n_r(0)\>,\nonumber\\
&D_{aa}^{\m\n}(x)=\<A^\m_a(x)\,A^\n_a(0)\>.
\end{align}
The correlators in the Schwinger-Keldysh basis are given by
\begin{align}\label{D_SK}
&D_{11}^{\m\n}(x)=\th(x^0)\<A^\m(x)A^\n(0)\>+\th(-x^0)\<A^\n(0)A^\m(x)\>,\no
&D_{22}^{\m\n}(x)=\th(-x^0)\<A^\m(x)A^\n(0)\>+\th(x^0)\<A^\n(0)A^\m(x)\>,\no
&D_{21}^{\m\n}(x)=\<A^\m(x)A^\n(0)\>,\no
&D_{12}^{\m\n}(x)=\<A^\n(0)A^\m(x)\>,
\end{align}
corresponding to time-ordered, anti-time-ordered, greater and lesser correlators respectively. From \eqref{ra_12}, we can relate correlators in the two basis as
\begin{align}
&D_{ra}^{\m\n}(x)=\frac{1}{2}\(D_{11}^{\m\n}(x)-D_{22}^{\m\n}(x)-D_{12}^{\m\n}(x)+D_{21}^{\m\n}(x)\),\no
&D_{ar}^{\m\n}(x)=\frac{1}{2}\(D_{11}^{\m\n}(x)-D_{22}^{\m\n}(x)+D_{12}^{\m\n}(x)-D_{21}^{\m\n}(x)\),\no
&D_{rr}^{\m\n}(x)=\frac{1}{4}\(D_{11}^{\m\n}(x)+D_{22}^{\m\n}(x)+D_{12}^{\m\n}(x)+D_{21}^{\m\n}(x)\),\no
&D_{aa}^{\m\n}(x)=D_{11}^{\m\n}(x)+D_{22}^{\m\n}(x)-D_{12}^{\m\n}(x)-D_{21}^{\m\n}(x).
\end{align}
Using the explicit representations in \eqref{D_SK} and $\th(x^0)+\th(-x^0)=1$, we easily find
\begin{align}\label{raar_RA}
&D_{ra}^{\m\n}(x)=\th(x^0)\<[A^\m(x),A^\n(0)]\>\equiv-i D_R^{\m\n}(x),\no
&D_{ar}^{\m\n}(x)=\th(-x^0)\<[A^\m(x),A^\n(0)]\>\equiv -i D_A^{\m\n}(x),\no
&D_{rr}^{\m\n}(x)=\frac{1}{2}\{A^\m(x),A^\n(0)\},\no
&D_{aa}^{\m\n}(x)=0.
\end{align}
with $D_{R}^{\m\n}(x)$ and $D_{A}^{\m\n}(x)$ being retarded and advanced correlators respectively.

In Schwinger-Keldysh basis, the vertices can be obtained from the interaction terms in the Lagrangian $J_1^\m(x)A_{1,\m}(x)-J_2^\m(x)A_{2,\m}(x)$. We can convert the current density $J_{1/2}$ to $J_{r/a}$ with a definition parallel to \eqref{ra_12} to arrive at the interaction terms $J_a^\m(x)A_{r,\m}(x)+J_r^\m(x)A_{a,\m}(x)$. The photon self-energy in $ra$-basis is simply the correlators of current density defined as follows
\begin{align}\label{Pi_def}
&\P_{ar}^{\m\n}(x)=\<J^\m_r(x)J^\n_a(0)\>,\nonumber\\
&\P_{ra}^{\m\n}(x)=\<J^\m_a(x)J^\n_r(0)\>,\nonumber\\
&\P_{aa}^{\m\n}(x)=\<J^\m_r(x)J^\n_r(0)\>.
\end{align}
Note that use the $ra$ labeling from the $A$ fields for $\P$ as is done conventionally. In particular $\P_{rr}$ instead of $\P_{aa}$ vanishes identically.

Now we focus on photon self-energy in charged magnetized plasma. The retarded self-energy is defined similar to \eqref{raar_RA} with $A\to J$. The results for neutral magnetized plasma in LLL approximation have been calculated using both field theory \cite{Fukushima:2011nu} and chiral kinetic theory \cite{Gao:2020ksg}. The inclusion of anti-symmetric part in charged magnetized plasma has also been made using field theory \cite{Fukushima:2019ugr} and chiral kinetic theory \cite{Lin:2021sjw,Yang:2021eoz}, with the results in momentum space quoted below
%
\begin{align}\label{rse}
&\P_{R}^{\a\b}=-\frac{e^3B}{2\p^{2}}\frac{q_{3}^{2} u^{\a}u^{\b}+q_{0}^{2}b^{\a}b^{\b}+q_{0}q_{3}u^{\{\a}b^{\b\}}}{\(q_{0}+i\e\)^2-q_{3}^{2}}+\frac{ie^2\m }{2\pi^{2}}\(q_{0} \e^{\a \b \r \s }+u^{[\a }\e^{\b]\l \r \s }q^T_{\l}\)u_{\r}b_{\s},
\end{align}
where we have defined $A^{\{\a}B^{\b\}}=A^{\a}B^{\b}+A^{\b}B^{\a}$ and $A^{[\a }B^{\b]}=A^{\a}B^{\b}-A^{\b}B^{\a}$. In \eqref{rse}, $B$ is the magnetic field and $\mu$ is the chemical potential for fermion number. For simplicity, we consider medium consisting of a single species of fermion carrying positive electric charge. $u^{\m}$ is fluid velocity and $b^{\m}$ is the direction of the magnetic field. $q_T^{\m}=b^{\m} \(q\cdot b\)+q^{\m}-u^{\m}\(q\cdot u\)$ corresponds to spatial components of $q$ perpendicular to $b^\m$.
The first term of \eqref{rse} is symmetric in indices with the pole coming from the chiral magnetic wave (CMW)) \cite{Kharzeev:2010gd} in the LLL approximation. The second term is anti-symmetric and purely imaginary. It comes from the Hall effect arising from the current along the drift velocity in charged plasma \cite{Lin:2021sjw,Yang:2021eoz}. This can be confirmed in field theory \cite{Fukushima:2019ugr} and in magnetohydrodynamics \cite{Hernandez:2017mch}. If we work in local rest frame of the plasma, and point the magnetic in $z$ direction so that $b^{\m}=(0,0,0,1)$ when $u^{\m}=(1,0,0,0)$. both \cite{Fukushima:2019ugr} and \cite{Hernandez:2017mch} give $\P_R^{xy}=\frac{i n_e}{B}q_0$ for $q_T=0$. In the LLL approximation, we can express the electric charge density $n_e$ in terms of electric chemical potential $\m_e=e\m$ and susceptibility $\c=\frac{eB}{2\p^2}$ as $n_e=\m_e \c=e\m\frac{eB}{2\p^2}$, which agrees with \eqref{rse}. The origin of the anti-symmetric component implies that \eqref{rse} is valid on a time scale longer than the relaxation time $\t_R$ such that Hall current can establish. This imposes an ultraviolet (UV) cutoff on $q_0$ with $q_0\lesssim 1/\t_R$. Beyond the cutoff, the Hall current has no time to develop leading to no anti-symmetric part in the self-energy.

\subsection{Electromagnetic wave in magnetized plasma}\label{subsec_maxwell}

We proceed to find the polarization modes for photon by solving the Maxwell equations in the magnetized plasma. We start with the Maxwell equations in coordinate space
\begin{align}
\(\pd^2\eta^{\m\n}-\pd^\m\pd^\n\)A_{\n,r}=j^{\m}_r=-i\int d^4y\P^{\m\n}_{ar}(x,y)A_{\n,r}.
\end{align}
Using \eqref{raar_RA} with $A\to J$, we can determine the following correlator in $ra$-basis
\begin{align}
\P^{\m\n}_{ar}(x)=-i \P_R^{\m\n}(x).
\end{align}
Working in momentum space and taking the Coulomb gauge $\nabla\cdot \vec{A}=0$ in local rest frame of the plasma, we can express the Maxwell equation as
\ba\label{MaxEq}
Q^2 A^{\m}-q_{0}A_{0}Q^{\m}-\Pi^{\m\n}_{R}A_{\n}=0.
\ea
The polarization modes for photon can be obtained from the solutions of $q_0^2$. For pedagogical purpose, we first solve \eqref{MaxEq} for neutral plasma $\m=0$, in which we obtain
\begin{align}\label{roots0}
&q_0^2=\tilde{B}+q^2+O(B^{-1}),\quad A_i=-\frac{A_0q_T^iq_0}{\tilde{B}},\;A_3=\frac{A_0q_T^2q_0}{\tilde{B}}\no
&q_0^2=q^2,\quad A_0=A_3=0,\;q_T^iA_i=0,
\end{align}
with $\tilde{B}=e^3B/2\p^2$ and $i=1,2$ labeling directions perpendicular to $b$. The first one is a gapped mode and the second one is lightlike\footnote{In the special case when $q_T=0$, the first mode disappears and the second mode becomes two degenerate ones, as photon does not feel the magnetic field. We are not interested in this trivial case.}.

Turning to the charged plasma, we can get three roots of $q_{0}^{2}$, corresponding to three polarization modes of the photon as follows
\ba\label{roots}
q_0^2=\tilde{B}+q^2,\no
q_0^2=\frac{1}{2}\(\tm^{2}+q_\perp^2+2q_3^2-\sqrt{4\tm^{2}q_3^2+\(q_\perp^2+\tm^{2}\)^2}\)\equiv x_{1}^2,\no
q_0^2=\frac{1}{2}\(\tm^{2}+q_\perp^2+2q_3^2+\sqrt{4\tm^{2}q_3^2+\(q_\perp^2+\tm^{2}\)^2}\)\equiv x_{2}^2.
\ea
with $\tm=e^2\m/ 2\pi^{2}$ and $q_\perp^2=q_1^2+q_2^2$. The first mode is the same gapped one as the neutral case. The second and third correspond to the space-like and the time-like low energy modes, respectively. The origin of the low energy modes is most clearly seen in the neutral limit where the two modes reduce to $q_0^2=q_3^2$ and $q_0^2=q^2$ respectively. The former corresponds to Landau damping, which arises from energy exchange between photon and LLL states. In the massless limit we consider, Landau damping appears as a pole instead of a cut \cite{Fukushima:2015wck}. The latter corresponds to photon dispersion in vacuum. The effect of finite density medium is to shift the two poles. The actual propagating modes are only the first and third ones in \eqref{roots}. One may expect to have three propagating modes rather than two due to collective motion in plasma \cite{Bellac:2011kqa}. Note that the self-energy \eqref{rse} contains no explicit temperature dependence, suggesting the medium is more like a Fermi sea rather than a plasma. It follows that the number of propagating modes matches that of the vacuum. We shall elaborate on this later.

Let us take a close look at the low energy modes in the phenomenologically motivated limit
\ba\label{mui}
\tm\gg q:\ \ x_{1}^{2}\approx\frac{q_{3}^{2}q^{2}}{\tm^{2}},\ \ \ \ \ \ x_{2}^{2}\approx\tm^{2}.
\ea
Now we argue that the mode $x_2$ actually lies outside the applicable region of \eqref{rse} by making an estimate of $\t_R$. Note that $\t_R$ is governed by dynamics perpendicular to the magnetic field, which is realized through $2\to2$ process \cite{Hattori:2016lqx}. We can estimate $\t_R$ as $\t_R^{-1}\sim e^4\m F((eB)^{1/2}/\m,T/\m)$ for some dimensionless function $F$. Regarding $\m\sim T$, we easily find $x_2\gg 1/\t_R$. Meanwhile, the condition $x_1\lesssim 1/\t_R$ for mode $x_1$ leads to $q\lesssim(\tm/\t_R)^{1/2}\sim e\tm\ll \tm$, which is consistent with our assumption in \eqref{mui}.
To gain further insights, we plug \eqref{roots} into \eqref{MaxEq} to solve for $A^{\m}$. In the same limit $\tm\gg q$, we obtain
\begin{align}\label{muli}
q_0^{2}&=x_{1}^{2}:\;\frac{A_1}{A_0}=\frac{i(q_2q+iq_1|q_3|)}{q_\perp^2q}\tm,\quad\frac{A_2}{A_0}=-\frac{i(q_1q-iq_2|q_3|)}{q_\perp^2q}\tm,\quad\frac{A_3}{A_0}=\frac{q_3}{|q_3|q}\tm.
\end{align}
The physical interpretation of this mode is most transparent if we focus on the regime $q_3\gg q_\perp$, that is a photon propagating almost along the magnetic field. We have then $\frac{A_1}{A_2}\simeq -i$ from \eqref{muli}. This is analogous to one of the circular polarization in vacuum, but with the dispersion modified by the charged medium. This parity breaking mode will play an important role in polarizing probe fermions just as a paramagnet polarizing an ordinary metal.

\subsection{Resummed photon propagator}\label{subsec_resum_photon}

In the previous section, we have obtained the photon polarization modes by solving the Maxwell equations. These modes contain pole and Landau damping (also a pole in massless limit) contributions to the spectral function of photon. In this section, we will derive the resummed photon propagator and extract the spectral function, from which we will find both pole and Landau damping contributions.

We start with the following bare photon propagators $D_{\m\n(0)}^{ar}$, $D_{\m\n(0)}^{ra}$ and $D_{\m\n(0)}^{rr}$ in Coulomb gauge in thermal equilibrium
\ba\label{bare_photon}
D_{\m\n(0)}^{ar}(Q)&=&\frac{i}{\(q_0-i\e \)^2-q^2}\(P_{\m\n}^T+\frac{Q^2 u_{\m}u_{\n}}{q^2}\),\no
D_{\m\n(0)}^{ra}(Q)&=&\frac{i}{\(q_0+i\e \)^2-q^2}\(P_{\m\n}^T+\frac{Q^2 u_{\m}u_{\n}}{q^2}\),\no
D_{\m\n(0)}^{rr}(Q)&=&2\p\ \sgn(q_{0})\ \d(Q^{2})\(\frac{1}{2}+f_{\g}(q_{0})\)\(P_{\m\n}^T+\frac{Q^2}{q^2}u_{\m}u_{\n}\).
\ea
with $\sgn(q_{0})$ being the sign function and $f_{\g}(q_{0})$ being the Bose-Einstein distribution function $f_{\g}(q_{0})=1/(\exp({q_{0}/T})-1)$. The structures $P_{\m\n}^T$ and $\frac{Q^2}{q^2}u_{\m}u_{\n}$ correspond to transverse and longitudinal components of the propagator respectively.
The transverse projection operator $P_{\m\n}^T$ is defined as $P_{\m\n}^T=P_{\m\n}-\frac{P_{\m\a}P_{\n\b}Q^\a Q^\b}{-Q^2+(Q\cdot u)^2}$ with $P_{\m\n}=u_\m u_\n-\eta_{\m\n}$ being the projection operator orthogonal to fluid velocity. In fluid's rest frame, we have
\ba
P_{00}^{T}=P_{0i}^{T}=P_{i0}^{T}=0,\no
P_{ij}^{T}=\d_{ij}-\frac{q_{i}q_{j}}{q^2}.
\ea
Using the definitions \eqref{D_def}, \eqref{Pi_def} and the couplings $J_a^\m(x)A_{r,\m}(x)+J_r^\m(x)A_{a,\m}(x)$, we may express the propagators up to first order in the self-energy as:
\ba\label{rmp}
\(
\begin{array}{c}
D^{rr}\ \ D^{ra}\no
D^{ar}\ \ \ 0\no
\end{array}
\)_{\m\n}
=
\(
\begin{array}{c}
D_{(0)}^{rr}\ \ D_{(0)}^{ra}\no
D_{(0)}^{ar} \ \ \ 0\no
\end{array}
\)_{\m\n}
-
\(
\begin{array}{c}
D_{(0)}^{rr}\ \ D_{(0)}^{ra}\no
D_{(0)}^{ar}\ \ \ 0\no
\end{array}
\)_{\m\a}
\(
\begin{array}{c}
0\ \ \ \  \P^{ra} \no
\P^{ar}\ \ \P^{aa}\no
\end{array}
\)^{\a\b}
\(
\begin{array}{c}
D_{(0)}^{rr}\ \ D_{(0)}^{ra} \no
D_{(0)}^{ar}\ \ \ 0\no
\end{array}
\)_{\b\n}.\no
\ea
By iteration, we deduce the resummed propagators satisfy the following equations
%
\ba\label{rmp}
\(
\begin{array}{c}
D^{rr}\ \ D^{ra}\no
D^{ar}\ \ \ 0\no
\end{array}
\)_{\m\n}
=
\(
\begin{array}{c}
D_{(0)}^{rr}\ \ D_{(0)}^{ra}\no
D_{(0)}^{ar} \ \ \ 0\no
\end{array}
\)_{\m\n}
-
\(
\begin{array}{c}
D_{(0)}^{rr}\ \ D_{(0)}^{ra}\no
D_{(0)}^{ar}\ \ \ 0\no
\end{array}
\)_{\m\a}
\(
\begin{array}{c}
0\ \ \ \  \P^{ra} \no
\P^{ar}\ \ \P^{aa}\no
\end{array}
\)^{\a\b}
\(
\begin{array}{c}
D^{rr}\ \ D^{ra} \no
D^{ar}\ \ \ 0\no
\end{array}
\)_{\b\n}.\no
\ea
The component form of the above reads
\ba\label{eqts}
D^{ra}_{\m\n}= D^{ra}_{\m\n(0)}-D^{ra}_{\m\a(0)}\P^{\a\b}_{ar}D_{\b\n}^{ra},\no
D^{ar}_{\m\n}= D^{ar}_{\m\n(0)}-D^{ar}_{\m\a(0)}\P^{\a\b}_{ra}D_{\b\n}^{ar},\no
D^{rr}_{\m\n}= D^{rr}_{\m\n(0)}-D^{ra}_{\m\a(0)}\P^{\a\b}_{ar}D_{\b\n}^{rr}-\(D_{\m\a(0)}^{rr}\P_{ra}^{\a\b}+D_{\m\a(0)}^{ra}\P_{aa}^{\a\b}\)D_{\b\n}^{ar}.
\ea
The resummed propagators can be solved by inverting the following matrix equations
\ba\label{reqts}
\(\d_{\a}{}^{\m}+D_{\a\b(0)}^{ra}\P_{ar}^{\b\m}\)D_{\m\n}^{ra}=D_{\a\n(0)}^{ra},\no
\(\d_{\a}{}^{\m}+D_{\a\b(0)}^{ar}\P_{ra}^{\b\m}\)D_{\m\n}^{ar}=D_{\a\n(0)}^{ar},\no
\(\d_{\a}{}^{\m}+D_{\a\r(0)}^{ra}\P_{ar}^{\r\m}\)D_{\m\n}^{rr}=\(D_{\a\n(0)}^{rr}-D_{\a\b(0)}^{rr}\P_{ra}^{\b\s}D_{\s\n}^{ar}-D_{\a\b(0)}^{ra}\P_{aa}^{\b\s}D_{\s\n}^{ar}\).
\ea
We first invert the first two equations to obtain $D_{\m\n}^{ra}(Q)$ and $D_{\m\n}^{ar}(Q)$, and then use the results to invert the last equation to obtain $D_{\m\n}^{rr}$. Note that our knowledge about the self-energy from the LLL approximation should be viewed as leading terms in the limit $B\to\infty$. It follows that we should also keep only the leading terms in the resulting resummed propagators, which gives the following results
%
\ba\label{srp}
D_{\m\n}^{ra}(Q)=\(\frac{1}{(q_{0}+i\e)^2-x_{1}^{2}}+\frac{1}{(q_{0}+i\e)^2-x_{2}^{2}}\)
\frac{A_{\m\n}(Q)\tm+S_{\m\n}(Q)}{\(q_{0}^{2}-x_{1}^{2}\)+\(q_{0}^{2}-x_{2}^{2}\)},\no
D_{\m\n}^{ar}(Q)=D_{\n\m}^{ra}(-Q),\no
D_{\m\n}^{rr}(Q)=-2i\p\ \sgn(q_{0})\(S_{\m\n}(Q)+A_{\m\n}(Q)\tm\)\(\frac{1}{2}+f_{\g}(q_{0})\)\(\frac{\d(q_{0}^{2}-x_{1}^{2})}{q_{0}^{2}-x_{2}^{2}}+\frac{\d(q_{0}^{2}-x_{2}^{2})}{q_{0}^{2}-x_{1}^{2}}\).
\ea
Here $A_{\m\n}$ and $S_{\m\n}$ are the anti-symmetric and symmetric tensors, defined respectively as
\ba\label{asmn}
A_{\m\n}(Q)&=&-\frac{q_{0}}{q^2}\(q_{0}u_{[\m}\e_{\n\l\r\s]}q_T^{\l}u^{\r}b^{\s}-q_{3}^{2}\e _{\m\n\r\s}u^{\r}b^{\s}+q_{3}b_{[\m }\e _{\n\l\r\s]}q_T^{\l}u^{\r}b^{\s}\),\no
S_{\m\n}(Q)&=&i\(-g_{\m\n}\(q_{0}^{2}-q_{3}^{2}\)-q_{3}^{2} u_{\m}u_{\n}-q_{0}^{2} b_{\m}b_{\n}-b_{\{\m}u_{\n\}}q_{0}q_{3}\)+\frac{i}{q^{2}}\(u_{\{\m}q_{\n\}}q_{0}^{3}+b_{\{\m}q_{\n\}}q_{0}^{2}q_{3}\)\no
&+&\frac{i}{q^4}q_{\m}q_{\n}\(q^{2}q_{3}^{2}-q_{0}^{2}\(q^{2}+q_{3}^{2}\)\).
\ea

Clearly the low energy modes found in Sec.~\ref{subsec_maxwell} are present as poles of $D_{\m\n}^{ra}(Q)$ and $D_{\m\n}^{ar}(Q)$. The gapped mode in Sec.~\ref{subsec_maxwell} is invisible after the limit $B\to\infty$ is taken in the resummed propagator. From the definition \eqref{D_def}, it is easy to show that $D_{\m\n}^{rr}(Q)$ is hermitian. This is indeed satisfied by the corresponding expression in \eqref{srp} with real symmetric and purely imaginary anti-symmetric components.

\section{Probe fermion in paramagnet}\label{pfpm}

We consider a probe fermion interacting with the medium. We choose an unmagnetized probe fermion. This is motivated by heavy ion phenomenology: with the quick decay of the magnetic field, the strange quarks produced at later stage are not spin polarized and can only interact with the medium. We shall consider high density limit $\tm\gg q$. In this case the medium is like a paramagnet, which is able to polarize probe fermion. We will corroborate the picture with calculations of damping rates of probe fermions. For simplicity, we take the probe fermion to be massless.

\subsection{Resummed fermion propagator and damping rate}
A probe fermion interacting with the medium will have a modified dispersion, with the damping rate given by imaginary part of the pole in the resummed retarded propagator.
The procedure of deriving resummed propagator is similar to section~\ref{subsec_resum_photon}. We start with the bare fermion propagators in $ra$-basis.
\ba\label{bfp}
S_{ar(0)}(P)=\frac{i\slashed{P}}{(p_{0}-i\e)^2-p^{2}}\no
S_{ra(0)}(P)=\frac{i\slashed{P}}{(p_{0}+i\e)^2-p^{2}}\no
S_{rr(0)}(P)=\(\frac{1}{2}-f_{e}(p_{0})\)2\p  \sgn(p_{0})\slashed{P}\d\(P^{2}\)
\ea
with $f_{e}$ being the Fermi-Dirac distribution function $f_e(p_0)=1/(\exp({p_0/T})+1)$. For the probe fermion, we set $f_{e}=0$. The resummation equation for retarded propagator is analogous to counterpart in \eqref{reqts}
\begin{align}\label{fermion_resum}
S_{ra}(P)=S_{ra(0)}(P)-S_{ra(0)}(P)\S_{ar}(P)S_{ra}(P).
\end{align}
The self-energy in \eqref{fermion_resum} is defined by the Fourier transform of the following
\begin{align}\label{Sigma_def}
\S_{ar}(x)=\<\h_r{\bar\h}_a\>,
\end{align}
\begin{align}\label{retarded_resum}
S_{ra}=\frac{i}{{\slashed P}+i\S_{ar}},
\end{align}
where we have dropped the $i\e$ assuming the self-energy $\S_{ar}$ already shifts the pole of $p_0$ from the real axis. Since both medium and probe fermions are chiral, the self-energy also preserves the chiral symmetry with the following decomposition
\begin{align}\label{Sigma_decomp}
\S_{ar}={\cal V}_\m \g^\m+{\cal A}_\m \g^5\g^\m.
\end{align}
with $\mathcal{A}_\mu$ and $\mathcal{V}_\mu$ being the vector and axial vector, respectively. The decoupling of left and right-handed components is manifest in chiral representation of Dirac matrices, with the following explicit denominator of \eqref{retarded_resum}
\begin{align}
{\slashed P}+i\S_{ar}=
\begin{pmatrix}
& \(P_\m+i{\cal V}_\m-i{\cal A}_\m\)\s^\m\\
\(P_\m+i{\cal V}_\m+i{\cal A}_\m\)\bar{\s}^\m&
\end{pmatrix}.
\end{align}
This allows us to treat left and right-handed components separately as
\begin{align}\label{S_resum_LR}
&S_{ra}^R=\frac{i}{\(P_\m+i{\cal V}_\m-i{\cal A}_\m\)\s^\m}=\frac{i\(P_\m+i{\cal V}_\m-i{\cal A}_\m\)\bar{\s}^\m}{(P+i{\cal V}-i{\cal A})^2}\no
&S_{ra}^L=\frac{i}{\(P_\m+i{\cal V}_\m+i{\cal A}_\m\)\bar{\s}^\m}=\frac{i\(P_\m+i{\cal V}_\m+i{\cal A}_\m\){\s}^\m}{(P+i{\cal V}+i{\cal A})^2}.
\end{align}
It is clear that the effect of self-energy is to shift the momenta of left and right-handed components respectively. The coefficient ${\cal A}$ encodes the splitting between left and right-handed components. At finite charge density, the medium is spin polarized. We suggest in Sec~\ref{subsec_maxwell} that the Landau damping mode is parity-breaking. Thus we expect splitting between left and right-handed components.

\begin{figure}[htbp]
	\centering
	\includegraphics[width=0.56\textwidth]{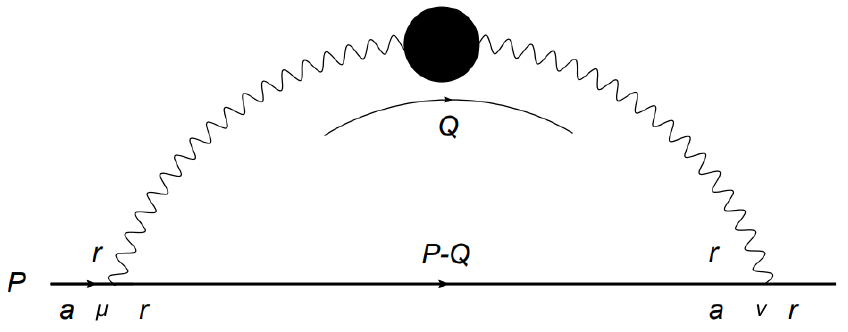}
	\vspace*{0.1cm}
	\caption{One-loop fermion self-energy $\S_{ar}$. Black solid circle represents the resummed photon propagator and the thick line indicates the probe fermion. The other diagram can be obtained by exchanging the $ra$-labeling of the photon and probe fermion in the loop. Its contribution does not give rise to damping because the probe fermion is not thermally populated.}\label{fm1}
\end{figure}

Now we present explicit calculation of the self-energy.
Fig.~\ref{fm1} shows one of the self-energy diagrams in $ra$-basis. The other diagram with $ra$ labelings of fermion and photon exchanged does not give rise to damping because temperature factor expected in damping rate cannot enter with the fermion being not thermally populated.
The corresponding self-energy contribution is given by\footnote{With our definition \eqref{Sigma_def}, the interaction vertex is $-e$ instead of $-ie$. The factor $i^2=-1$ appears in the resummation equation \eqref{fermion_resum}.}
\ba\label{sra0}
\S_{ar}(P)=e^{2}\int \frac{d^{4}Q}{(2\p)^{4}}\g^{\m}S_{ra(0)}(P-Q)\g^{\n}D_{\m\n}^{rr}(Q).
\ea

\subsection{Damping in paramagnet}

Now we evaluate the self-energy in the high density limit $\tm\gg q$. We have shown in Sec~\ref{subsec_maxwell} that only the Landau damping mode survives in this limit. We then evaluate the integrals with \eqref{bfp} and \eqref{srp} taking contribution from $q_0^2=x_1^2$ only. We calculate separately anti-symmetric and symmetric contributions to be denoted as $\S_{ar}^A$ and $\S_{ar}^S$. The anti-symmetric contribution reads
\ba\label{sra1}
\S_{ar}^{A}(P)&=&\frac{e^{2}}{(2\p)^{3}}\int \frac{d^{4}Q\ \sgn(q_{0})}{(P-Q)^2+i\e(p_{0}-q_{0})}\(\frac{1}{2}+f_{\g}(q_{0})\)\(\frac{\d(q_{0}^{2}-x_{1}^{2})}{q_{0}^{2}-x_{2}^{2}}+\frac{\d(q_{0}^{2}-x_{2}^{2})}{q_{0}^{2}-x_{1}^{2}}\)\no
&\times&\g^{\m}(\slashed{P}-\slashed{Q})\g^{\n}A_{\m\n}(Q)\tm
\ea
We first deal with $\g^{\m}(\slashed{P}-\slashed{Q})\g^{\n}A_{\m\n}(Q)$ by using the following relation
\begin{align}
\g^{\m}\g^{a}\g^{\n}=g^{\m\a}\g^{\n}-g^{\m\n}\g^{\a}+g^{\a\n}\g^{\m}-i\e^{\m\a\n\b}\g^{5}\g_{\b}.
\end{align}
Only the last anti-symmetric term contributes when contracted with $A_{\m\n}$, giving
\ba
\g^{\m}(\slashed{P}-\slashed{Q})\g^{\n}A_{\m\n}(Q)=-i(P-Q)_{\a}\e^{\m\a\n\b}\g^{5}\g_{\b}A_{\m\n}(Q)
=\frac{2i}{q^2}\(q_{0}^2 f_{1}+q_{0} f_{2}\)
\ea
with
\ba\label{f12}
f_{1}=\({\bf p}_\perp\cdot {\bf q}_\perp-q^{2}\)\g^{5}\g^{3}-p_{3}\g^{5}{\bf q}_\perp\cdot{\boldsymbol \g}_\perp\no
f_{2}=p_{0}q_{3}^{2}\g^{5}\g^{3}+p_{0}q_{3}\g^{5}{\bf q}_\perp\cdot{\boldsymbol  \g}_\perp+q_{3}\g^{5}\g^{0}(q^2-{\bf p}\cdot{\bf q}),
\ea
being the coefficients of even and odd powers of $q_0$. The perpendicular vectors are defined as ${\bf p}_\perp=(p_1,p_2)$ and similarly for ${\bf q}_\perp$ and ${\boldsymbol \g}_\perp$.

We proceed by making several approximations: firstly the self-energy induces only a small correction to the dispersion, so for the purpose of finding damping rate of on-shell probe fermion we may set $P^2=0$; secondly the Landau damping mode is nearly static, allowing us to approximate $\frac{1}{2}+f_{\g}(Q)\simeq\frac{T}{q_{0}}$; thirdly combining the on-shell condition and $q_0\ll q\ll p_0$, we approximate the denominator of fermion propagator as
%
\ba
\frac{1}{(P-Q)^2+i\e(p_{0}-q_{0})}\simeq \frac{1}{2}\frac{1}{\(\bold{p}\cdot\bold{q}+i\e p_{0}\)},
\ea
dropping $Q^2\ll 2P\cdot Q$ and $q_0p_0$.
Then, \eqref{sra1} can be written as
\ba\label{srat}
\S_{ar}^{A}(P)=-\frac{ie^{2}T}{(2\p)^{3}\tm}\int \frac{d^{3}q}{q^{2}\(\bold{p}\cdot\bold{q}+i\e p_{0}\)}\int \frac{d q_{0}}{q_{0}}\d\(q_{0}^{2}-\frac{q_{3}^{2}q^{2}}{\tm^{2}}\)\sgn(q_{0})\(q_{0}^2 f_{1}+q_{0} f_{2}\).
\ea
We proceed with the integral of $q_{0}$ first. Since $f_{1}$ and $f_{2}$ are independent of $q_{0}$, the integral receives contribution from integrand even in $q_{0}$ as
\ba
\int \frac{d q_{0}}{q_{0}}\d\(q_{0}^{2}-\frac{q_{3}^{2}q^{2}}{\tm^{2}}\)\e(q_{0})\(q_{0}^2 f_{1}+q_{0} f_{2}\)=f_{1}\int d q_{0}\ \d\(q_{0}^{2}-\frac{q_{3}^{2}q^{2}}{\tm^{2}}\) q_{0}\ \sgn(q_{0}) =f_{1}.
\ea
The remaining integrals are evaluated using the residue theorem. The details can be found in Appendix. We quote the final results here. To be specific, we take $p_{0}>0$ to arrive at the following results
\ba\label{sarmm}
\S_{ar}^{A}(P)&=&-i c_{1}\g^{5}\g^{3}+i c_{2}\g^{5}\bold{p}_{\perp}\cdot\boldsymbol{\g}_\perp+c_{3}\g^{5}\g^{3},
\ea
with
\ba
c_{1}=\frac{e^{2}T\quv}{4\p\tm}\(1-\frac{|p_{3}|}{p}\),\ \
c_{2}=\frac{e^{2}T\quv}{4\p\tm}\(1-\frac{|p_{3}|}{p}\)\frac{ p_{3}}{p_{\perp}^2},\ \
c_{3}=\frac{e^{2}T \quv^{2}}{8\p\tm |p_{3}|}.\ \
\ea
$\quv$ is the UV cutoff for $q$, which can be taken as $\quv\sim e\tm$ based on the discussion below \eqref{mui}.

%

The calculation of the symmetric contribution proceeds similarly. We simply quote the final result, collecting details in Appendix. For $p_0>0$ we have
\begin{align}\label{asarmm}
\S_{ar}^S=\g^0d_1+i{\bf p}_\perp\cdot{\boldsymbol \g}_\perp d_2+i\g^3d_3,
\end{align}
with
\begin{align}
d_1=\frac{e^2T}{4\p}\frac{p_0\ln\frac{\quv}{\qir}}{p},\quad d_2=\frac{e^2T}{4\p}\frac{\quv}{p_\perp^2}\(1-\frac{|p_3|}{p}\),\quad
d_3=\frac{e^2T}{4\p}\frac{\quv}{p}\sgn(p_3).
\end{align}
$\qir$ is the infrared (IR) cutoff of $q_\perp$. We will discuss the meaning of the IR cutoff shortly.

Now we can take $\S_{ar}=\S_{ar}^A+\S_{ar}^S$ and compare with \eqref{Sigma_decomp} and \eqref{S_resum_LR} to obtain damping rates of left and right-handed components respectively. We find it more instructive to obtain the contributions to damping rate from $\S_{ar}^A$ and $\S_{ar}^S$ respectively. In fact, if we keep linear order in $\S_{ar}^A$ and $\S_{ar}^S$, the corresponding shifts of the poles from the vacuum counterpart are additive. The imaginary part of the shift gives the damping rate.
We first consider contribution from $\S_{ar}^A$. Using \eqref{Sigma_decomp} and \eqref{S_resum_LR}, we easily find the poles given by
\ba
L: p_{0}\simeq p+\frac{c_2p_\perp^2}{p}-\frac{c_1p_3}{p}-\frac{ic_3p_3}{p}=p-\frac{ic_3p_3}{p},\no
R: p_{0}\simeq p-\frac{c_2p_\perp^2}{p}+\frac{c_1p_3}{p}+\frac{ic_3p_3}{p}=p+\frac{ic_3p_3}{p}.
\ea
We can see $\S_{ar}^A$ causes the shifts of poles with opposite sign for both real and imaginary parts. The real part would lead to the chiral shift discussed in \cite{Gorbar:2009bm}, but vanishes when explicit expressions for $c_1$ and $c_2$ are used. The imaginary part gives the following damping rates 
\ba\label{damping_A}
\G_L\simeq \frac{c_3p_3}{p}=\frac{e^{2}T \quv^{2}}{8\p\tm p}\sgn(p_3),\no
\G_R\simeq -\frac{c_3p_3}{p}=-\frac{e^{2}T \quv^{2}}{8\p\tm p}\sgn(p_3).
\ea
The cases with $\G<0$ are unstable. These include right-handed component with $p_3>0$ and left-handed component $p_3<0$. The implication is interesting: Due to spin-momentum locking, both cases have a positive spin component along the direction of the paramagnet. Interaction with the paramagnet tends to polarize the probe fermion by amplifying these modes. In contrast, left-handed component with $p_3>0$ and right-handed component with $p_3<0$ have $\G>0$. They both have a negative spin component along the direction of the paramagnet and are damped out. This provides a mechanism to polarize the probe fermion.

Now we turn to the symmetric contribution. This contribution leads to identical shifts for the left and right-handed components:
\begin{align}
p_0\simeq p+\frac{p_\perp^2d_2}{p}+\frac{p_3d_3}{p}-id_1=p+\frac{e^2T}{4\p}\frac{\quv}{p}-id_1.
\end{align}
The corresponding damping rate is given by
\begin{align}
\G=d_1=\frac{e^2T}{4\p}\frac{p_0\ln\frac{\quv}{\qir}}{p}.
\end{align}
It depends on both UV and IR cutoffs. While the UV cutoff is set by $\quv\sim e\tm$, the IR cutoff is fictitious. The appearance of the logarithmic divergence is not new. By using the resummed photon propagator, we have soften the $\int{dq}/{q^3}$ IR divergence typical in Coulomb scattering into a logarithmic one. The persistence of logarithmic divergence indicates the corresponding damping is non-exponential in time, which can be obtained with more sophisticated resummation \cite{Blaizot:1996hd}. Since we are mainly concerned with the splitting in damping rates, we shall not attempt further resummation as in \cite{Blaizot:1996hd}.

Combining the contributions from anti-symmetric and symmetric parts, we obtain a slightly modified picture: probe fermion interacting with the medium will generically be damped. This is because the damping rate from the symmetric contribution is parametrically larger than the counterpart from the anti-symmetric contribution: $d_1\sim e^2T\gg\frac{c_3|p_3|}{p}\sim \frac{e^6T\m}{p}$ using $\quv\sim e\tm$. However, with the medium being like a paramagnet, modes with positive/negative spin component along the direction of the paramagnet have smaller/larger damping rate, thus interaction tends to polarize the probe fermion. This occurs at a time scale $t\sim \D\G^{-1}\sim \frac{p}{e^6T\m}$. One may worry that at this time scale, the probe fermion has been damped out completely because of the hierarchy $d_1\gg\frac{c_3|p_3|}{p}$. This can still have physical consequence. If the probe fermion is continuously produced in the medium, the number density can maintain despite of damping by the medium, but the polarization mechanism from the splitting of damping rates always works. We will extend the analysis to QGP case in the next section, where we will see the splitting of damping rates can be parametrically enhanced, making the polarization dynamics more efficient.

\section{Probe quark in paramagnet of QGP}\label{sec-QGP}

Now we extend the analysis to probe quark in charged QGP. A new feature in this case is that gluon self-energy receives an additional contribution from gluon self-interaction. It follows that the dispersions we obtain from solving Maxwell equations no longer apply. We will identify low energy modes by finding the resummed gluon propagator and use it to calculate the splitting of damping rates for probe quark.

\subsection{Gluon propagator in charged QGP}

We follow the procedure in Sec.~\ref{subsec_resum_photon}. The gluon bare propagator in Coulomb gauge is the same as \eqref{bare_photon} except for additional color structures
\begin{align}
&D_{\m\n(0)}^{AB,ar}=\d^{AB}D_{\m\n(0)}^{ar},\no
&D_{\m\n(0)}^{AB,ra}=\d^{AB}D_{\m\n(0)}^{ra},\no
&D_{\m\n(0)}^{AB,rr}=\d^{AB}D_{\m\n(0)}^{rr}.
\end{align}
We have used capital letters for color indices and the color structure is diagonal $\d^{AB}$.
The gluon self-energy is given by
\begin{align}\label{gluon_Pi}
&\P_{R}^{\m\n,AB}=\bigg[-\frac{g^2eB}{2\p^{2}}\frac{q_{3}^{2} u^{\m}u^{\n}+q_{0}^{2}b^{\m}b^{\n}+q_{0}q_{3}u^{\{\m}b^{\n\}}}{\(q_{0}+i\e\)^2-q_{3}^{2}}+\frac{ig^2 }{2\pi^{2}}\frac{\m}{2}\(q_{0} \e^{\m \n \r \s }+u^{[\m }\e^{\n]\l \r \s }q^T_{\l}\)u_{\r}b_{\s}\no
&-P^{\m\n}_T\P_T-P^{\m\n}_L\P_L\bigg]\d^{AB},
\end{align}
with $\P_{T/L}$ being the transverse/longitudinal components from gluon loop. The explicit expressions in the hard thermal loop (HTL) regime are as follows
\begin{align}\label{HTL_Pi}
&\P_T=m^2\(x^2+(1-x^2)x Q_0(x)\),\no
&\P_L=-2m^2(x^2-1)\(1-x Q_0(x)\),
\end{align}
where $m^2=\frac{1}{6}N_cg^2T^2$ is the thermal mass and $N_c$ is the number of colors. The Legendre function $Q_0$ is defined as $Q_0(x)=\frac{1}{2}\ln\lvert\frac{x+1}{x-1}\rvert-\frac{i\p}{2}\th(1-x^2)$. The symmetric components of \eqref{gluon_Pi} have been extensively discussed in \cite{Hattori:2017xoo}. The anti-symmetric component is obtained by a straightforward generalization of the calculations in \cite{Fukushima:2019ugr} for a single species of quark carrying positive electric charge $q_f>0$, with $\m$ being chemical potential for quark number density. The overall factor $\frac{1}{2}$ comes from color trace in the fundamental representation $\tr [t^A t^B]=\frac{1}{2}\d^{AB}$. The physical interpretation is the chromo-Hall effect. Imagine applying a chromo-electric field in color direction $A$ perpendicular to the magnetic field. The quarks carrying both electric charge $q_f$ and effective chromo charge $\bar{g}$ will develop a drift velocity $v^A=\frac{\bar{g}E^A}{q_fB}$ where the chromo-electric force and ordinary Lorentz force reaches a balance. This gives rise to a chromo current along the drift velocity
\begin{align}
J^A= \bar{g}\r v^A=\frac{\bar{g}^2E^A}{q_fB}\c\m=\bar{g}^2E^A\m,
\end{align}
where we have used $\c=q_fB$. To arrive at \eqref{gluon_Pi}, we need to fix the effective chromo charge. This is most easily done in double line basis for color \cite{Hidaka:2009hs}, in which the gluon color index is represented as $A=ij$ and quark color indices are represented by $i$ and $j$. The color matrices in fundamental representation are given by
\begin{align}
t^{ij}_{kl}=\frac{1}{\sqrt{2}}\(\d^i_k\d^j_l-\frac{1}{N_c}\d^{ij}\d_{kl}\).
\end{align}
It is most easily understood in the large $N_c$ limit, in which the color indices of gluons and quarks are locked. Naturally the corresponding quarks lead to chromo current in the same color direction as the chromo-electric field with the effective charge $\bar{g}=\frac{1}{\sqrt{2}}g$, thus the factor $\frac{1}{2}$ is perfectly accounted for.

Since the color structure is trivial in both bare propagator and self-energy, we can simply ignore it and then use \eqref{eqts} to obtain the resummed gluon propagator. We assume the following hierarchy: $eB\gg T^2$, $eB\gg \m q$ and expand to leading order in $B^{-1}$. The resulting resummed propagator contains both symmetric and anti-symmetric parts. The symmetric part exists in the absence of $\m$. It does not lead to splitting of damping rate so we do not keep track of. The anti-symmetric part is given by (suppressing the color structure)
\begin{align}\label{DA_gluon}
&D_{\m\n}^{ra,A}(Q)=-q^2Q^2\bigg[-q^6(q_0^2-q_3^2)+q_0^2q_3^2(q_0^2-\P_T)(\P_T-\P_L)+q^4\big[2q_0^4+2q_3^2\P_T-q_0^2(2q_3^2+\bmu^2 \no
&+\P_T+\P_L)\big] +q^2\big[-q_0^6+q_3^2\P_T^2+q_0^4(q_3^2+\bmu^2+\P_T+\P_L)+q_0^2(q_3^2(\P_L-3\P_T)-\P_T\P_L)\big]\bigg]^{-1}
A_{\m\n}\bmu,\no
&D_{\m\n}^{ar,A}(Q)=D_{\n\m}^{ra,A}(-Q),\no
&D_{\m\n}^{rr,A}(Q)=\(D_{\m\n}^{ra,A}(Q)-D_{\m\n}^{ar,A}(Q)\)\(\frac{1}{2}+f_g(q_0)\),
\end{align}
with $\bmu=\frac{\bar{g}^2\m}{2\p^2}$ as analog of $\tm$. It should be understood that $q_0\to q_0+i\e$ is needed in the first explicit expression.
The resummed $rr$-propagator is constructed using spectral representation, with $D_{\m\n}^{ra,A}(Q)-D_{\m\n}^{ar,A}(Q)$ giving the anti-symmetric part of the spectral function and $f_g$ being Bose-Einstein distribution for gluons. It is worth noting that \eqref{DA_gluon} and \eqref{srp} share the same Lorentz structure, but very different spectral function. The spectral function arises from collective excitations  as well as Landau dampings. The Landau damping can give both pole and cut contributions from gluon-quark (in the LLL state) scatterings and gluon-gluon scatterings respectively.

To simplify the analysis, we consider two limits where either one of them dominates: $\bmu^2\gg \P_{T/L}$ (high density limit) and $\bmu^2\ll \P_{T/L}$ (low density limit). Note that $\P_{T/L}\sim g^2 T^2$. The former limit requires parametrically large $\m$ or small $T$: $g^2\m\gg T$. In this case, we may ignore $\P_{T/L}$ and the $rr$-propagator reduces to the QED counterpart \eqref{srp} with the substitution $\tm\to\bmu$:
\begin{align}\label{DA_gluon_mu}
D_{\m\n}^{rr,A}(Q)=-2i\p\ \sgn(q_{0})\(\frac{1}{2}+f_{g}(q_{0})\)\(\frac{\d(q_{0}^{2}-\bar{x}_{1}^{2})}{q_{0}^{2}-\bar{x}_{2}^{2}}+\frac{\d(q_{0}^{2}-\bar{x}_{2}^{2})}{q_{0}^{2}-\bar{x}_{1}^{2}}\)A_{\m\n}(Q)\bmu.
\end{align}
We have used $\bar{x}_{1,2}$ to denote $x_{1,2}(\tm\to\bmu)$. Like in the QED case, the spectral function contains Landau damping poles and lightlike mode, both modified by density.
The latter limit is close to the usual magnetized plasma limit, with the density leads to the following $rr$-propagator
\begin{align}\label{DA_gluon_T}
D_{\m\n}^{rr,A}(Q)=2i\text{Im}\[\frac{Q^2q^2}{(Q^2-\P_T)\(q^2Q^2(q_0^3-q_3^2)-Q^2q_3^2\P_T-q_0^2q_\perp^2\P_L\)}\]\(\frac{1}{2}+f_g(q_0)\)A_{\m\n}\bmu.
\end{align}
the spectral function contains two poles and cut. The poles are located at
\begin{align}
Q^2-\P_T=0,\quad q^2Q^2(q_0^3-q_3^2)-Q^2q_3^2\P_T-q_0^2q_\perp^2\P_L=0.
\end{align}
Not surprisingly they correspond to the transverse mode and mixed mode in HTL regime in the absence of $\m$ \cite{Hattori:2017xoo}. The location of the cut is at $Q^2<0$, from Landau damping. Although the anti-symmetric part inherits most spectral features from the symmetric part, there is one difference: the transverse mode and mixed mode are decoupled in the symmetric part, but are coupled in the anti-symmetric part in the form of product in \eqref{DA_gluon_T}.

\subsection{Damping rate of probe quark}

Now we can proceed to calculate the splitting of damping rates from anti-symmetric part of gluon propagator. Similar to \eqref{sra0}, we have for the quark self-energy
\begin{align}\label{quark_A}
\S_{ar}(P)=\frac{N_c^2-1}{2N_c}g^{2}\int \frac{d^{4}Q}{(2\p)^{4}}\g^{\m}S_{ra(0)}(P-Q)\g^{\n}D_{\m\n}^{rr}(Q),
\end{align}
where the overall factor comes from $t^At^A=\frac{N_c^2-1}{2N_c}$. The integration of $Q$ should be bounded by the applicable region of chromo-Hall dynamics, which will be fixed by $q_0\lesssim \t_R^{-1}$.

We first calculate the damping rate in the high density limit $\bmu^2\gg \P_{T/L}$. Using \eqref{DA_gluon_mu}, \eqref{quark_A} and \eqref{bfp}, we have
\begin{align}
&\S_{ar}^{A}(P)=\frac{N_c^2-1}{2N_c}\frac{g^{2}}{(2\p)^{3}}\int \frac{d^{4}Q\ \sgn(q_{0})}{(P-Q)^2+i\e(p_{0}-q_{0})}\(\frac{1}{2}+f_{g}(q_{0})\)\(\frac{\d(q_{0}^{2}-\bar{x}_{1}^{2})}{q_{0}^{2}-\bar{x}_{2}^{2}}+\frac{\d(q_{0}^{2}-\bar{x}_{2}^{2})}{q_{0}^{2}-\bar{x}_{1}^{2}}\)\no
&\times\g^{\m}(\slashed{P}-\slashed{Q})\g^{\n}A_{\m\n}(Q)\bmu
\end{align}
The gamma matrices are evaluated in the same way as before
\begin{align}
\g^{\m}(\slashed{P}-\slashed{Q})\g^{\n}A_{\m\n}(Q)
=\frac{2i}{q^2}\(q_{0}^2 f_{1}+q_{0} f_{2}\),
\end{align}
with $f_1$ and $f_2$ taking the schematic forms $c_\m\g^5\g^\m$ and $c_\m$ being real functions of $P$ and $Q$. We make the following observation: the damping rate arises from purely imaginary shift of momentum. This corresponds to real part of the coefficients of $\g^5\g^\m$ in $\S_{ar}^A$. It is only possible when the $i\e$ prescription is invoked in the integral. It amounts to keeping the real part of the following
\begin{align}
\text{Re}\frac{i}{(P-Q)^2+i\sgn(p_0-q_0)}=\p\d((P-Q)^2)\sgn(p_0-q_0).
\end{align}
Taking $P^2=0$ as before and $p_0>0$, we have the $\d((P-Q)^2)\simeq \d(2P\cdot Q)$. The Dirac delta function is non-vanishing for spacelike $Q$ only, which comes from the mode $q_0^2=\bar{x}_1^2$. This is the density modified Landau damping pole. If we further estimate the relaxation time in high density limit $\t_R^{-1}\sim g^4\m G((eB)^{1/2}/\m)$ for some dimensionless function $G$. The condition $x_1\lesssim 1/\t_R$ leads to $q\lesssim g\bmu\ll\bmu$ similar to the QED case. Also, the mode becomes almost static as $x_1^2\simeq \frac{q_3^2q^2}{\bmu^2}\ll q^2$. The remaining analysis is parallel to the QED case. We can easily obtain the damping rate by the substitution $\tm\to\bmu$ in \eqref{damping_A}
\begin{align}
&\G_L\simeq\frac{N_c^2-1}{2N_c}\frac{g^2T\quv^2}{8\p\bmu p}\sgn(p_3),\no
&\G_R\simeq-\frac{N_c^2-1}{2N_c}\frac{g^2T\quv^2}{8\p\bmu p}\sgn(p_3),
\end{align}
with $\quv\sim g\bmu$.
Clearly QGP in this limit behaves like a paramagnet, which amplifies/damps mode with positive/negative spin component along the magnetic field respectively.

Now we move on to the limit $\bmu^{2}\ll\P_{T/L}$. Using \eqref{DA_gluon_T}, \eqref{quark_A} and \eqref{bfp}, we obtain the following representation
\begin{align}
&\S_{ar}^A(P)=\frac{N_c^2-1}{2N_c}g^2\int\frac{d^4Q}{(2\p)^4}\g^\m\frac{i({\slashed P}-{\slashed Q})}{(P-Q)^2+i\e(p_0-q_0)}\g^\n \(\frac{1}{2}+f_g(q_0)\)\no &\times 2i\text{Im}\[\frac{Q^2q^2}{(Q^2-\P_T)\(q^2Q^2(q_0^3-q_3^2)-Q^2q_3^2\P_T-q_0^2q_\perp^2\P_L\)}\]A_{\m\n}\bmu.
\end{align}
As reasoned above, only spacelike $Q$ in the spectral function contributes to the damping rate. In this case, it corresponds to the Landau damping cut, from which we obtain
\begin{align}
&\text{Re}\S_{ar}^A(P)=\frac{N_c^2-1}{2N_c}g^2\int\frac{d^3q}{(2\p)^4}\frac{\p}{2p} \frac{T}{q_0}\(\frac{2i\bmu}{q^2}\)\(q_{0}^2 f_{1}+q_{0} f_{2}\)\no &\times 2i\,\text{Im}\[\frac{Q^2q^2}{(Q^2-\P_T)\(q^2Q^2(q_0^3-q_3^2)-Q^2q_3^2\P_T-q_0^2q_\perp^2\P_L\)}\]\vert_{q_0=\hat{{\bf p}}\cdot {\bf q}}.
\end{align}
We can further simplify the integral by noting that $\text{Im}\[\frac{Q^2q^2}{(Q^2-\P_T)\(q^2Q^2(q_0^2-q_3^2)-Q^2q_3^2\P_T-q_0^2q_\perp^2\P_L\)}\]$ is odd in $q_0$ thus also odd under ${\bf q}\to-{\bf q}$. To have an integrand even under ${\bf q}\to-{\bf q}$, we can just keep the following terms in $f_1$ and $f_2$
\begin{align}\label{f12_part}
f_1=-q^2\g^5\g^3,\quad f_2=q_3q^2\g^5\g^0.
\end{align}
We then parameterize the quark self-energy as
\begin{align}\label{ReSigmaA}
\text{Re}\S_{ar}^A(P)\equiv\frac{N_c^2-1}{2N_c}g^2\int\frac{dq}{(2\p)^4}\frac{\p T\bmu}{p}\(h_1\g^5\g^3+h_2\g^5\g^0\).
\end{align}
By rotational invariance, $h_1$ and $h_2$ are even and odd functions of $\hat{p}_3$ respectively. Their precise forms can only be obtained numerically. We use the following parameterization of $q$
\begin{align}
{\bf q}= q\cos\a\hat{  p}+q\sin\a\cos\b\frac{\hat{ b}-\cos\g\,\hat{ p}}{\sin\g}+q\sin\a\sin\b\frac{\hat{ b}\times\hat{ p}}{\sin\g}.
\end{align}
We have chosen $\hat{ p}$ as the $z$-axis and the plane spanned by $\hat{ p}$ and $\hat{ b}$ as the $z-x$ plane. $\g$ denotes the angle between $\hat{ p}$ and $\hat{ b}$ with $\cos\g=\hat{ p}\cdot\hat{ b}$. We have then $\hat{ p}\cdot \hat{ q}=\cos\a$ and $d^3q=q^2dq d\cos\a d\b$. The angular integration is performed numerically to obtain $h_{1,2}$.

\begin{figure}[htbp]
	\centering
	\includegraphics[width=0.56\textwidth]{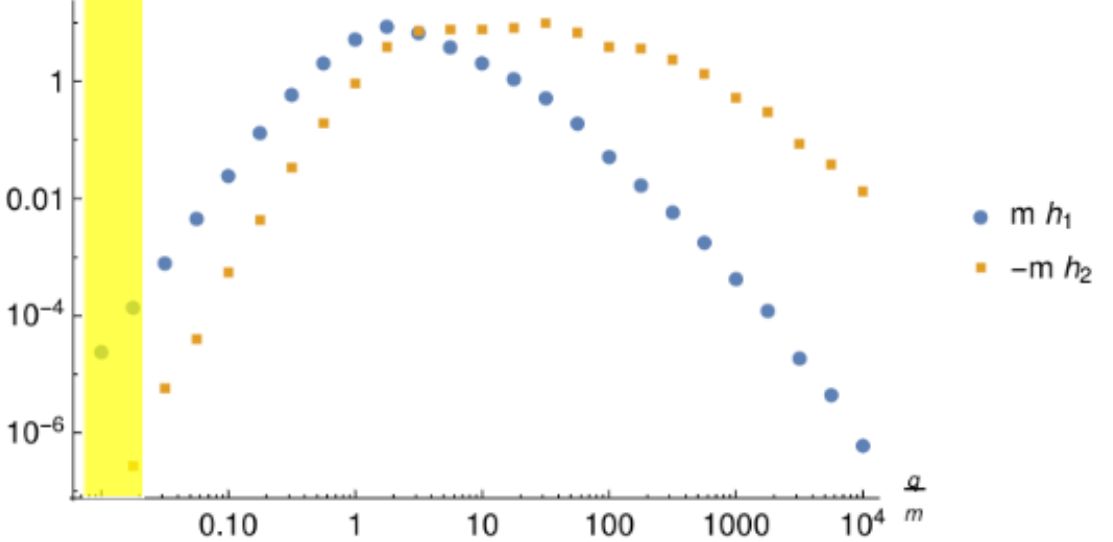}
	\vspace*{0.1cm}
	\caption{$q$-dependence of $h_1$ (disk) and $h_2$ (square) for $\hat{p}\cdot\hat{b}=\cos\g=\frac{1}{3}$. Both are finite in the IR and UV. The shape of the $q$-dependence remains qualitatively the same for a wide range of $\hat{p}\cdot\hat{b}$. With $\quv\ll m$, the integration range is schematically indicated by the yellow band. It follows that the integration of $h_1$ dominates the counterpart of $h_2$.}\label{h12-q}
\end{figure}

The $q$-dependence is of particular interest. It has been shown that the dynamical screening crucial for damping rate is the same as the case without magnetic field in the IR limit \cite{Hattori:2017xoo}. It follows that damping rate from symmetric contribution contains logarithmic divergence \cite{Braaten:1991we}. One may expect similar logarithmic divergence in the splitting of damping rate from anti-symmetric contribution. It turns out that this is not the case. Fig.~\ref{h12-q} shows the $q$-dependence of $h_{1,2}$ for a generic $\cos\g$. Both $h_{1}$ and $h_2$ are IR safe. In the UV $h_2$ decays more slowly than $h_1$.
Let us we define the $q$-integrated quantities as
\begin{align}
\text{Re}\S_{ar}^A(P)=H_1\g^5\g^3-\e(p_3)H_2\g^5\g^0,
\end{align}
with
\begin{align}
H_1=\frac{N_c^2-1}{2N_c}g^2\frac{\p T\bmu}{(2\p)^4p}\int_0^{\quv} dq h_1,\quad H_2=\frac{N_c^2-1}{2N_c}g^2\frac{\p T\bmu}{(2\p)^4p}\int_0^{\quv} dq |h_2|.
\end{align}
We have taken into account the signs of $h_1$ and $h_2$ (note that the latter is an odd function of $\hat{p}_3$) such that both $H_1$ and $H_2$ are positive. This leads to the following dispersions:
\ba\label{disp_LR}
L: p_{0}\simeq p-\frac{ip_3H_1}{p}+i \sgn(p_3)H_2,\no
R: p_{0}\simeq p+\frac{ip_3H_1}{p}-i \sgn(p_3)H_2,
\ea
with the following damping rate
\ba\label{Gamma_LR}
\G_L\simeq -\sgn(p_3)\(H_2-\frac{|p_3|H_1}{p}\),\no
\G_R\simeq \sgn(p_3)\(H_2-\frac{|p_3|H_1}{p}\).
\ea
To determine which mode is unstable, we need to compare $H_2$ with $\frac{|p_3|H_1}{p}$, which depends on the choice of $\quv$. In the magnetized plasma limit with the cut contribution, we may estimate $q\sim q_0\sim \t_R^{-1}\sim g^4T$, thus $\quv\sim g^4T\ll m$. From Fig.~\ref{h12-q}, we see $h_1$ dominates over $h_2$ in the range of integration schematically indicated by the yellow band, so we may consider the $H_1$ term in \eqref{Gamma_LR} only. It follows that \eqref{Gamma_LR} has the same structure as \eqref{damping_A} so that the previous reasoning applies: the right-handed mode with $p_3>0$ and left-handed mode with $p_3<0$ are amplified with respect to their chiral partners. This is just the amplification of the mode with a positive spin component along the magnetic field, which provides a mechanism to polarize the probe quark by the QGP paramagnet.

We are ready to propose the following picture for polarization dynamics in heavy ion collisions: the initial strong magnetic field first polarizes the spin of light quarks in the QGP. At not very high energy, the QGP carries finite baryon density. Due to mismatch of charges of up and down quarks, the medium is also electrically charged, and thus can be treated as a magnet. We have analyzed the high density or low temperature limit and magnetized plasma limit. In both cases, the QGP behaves as paramagnet. The initial magnetic field decays quickly so cannot affect the strange quarks produced in late stage of heavy ion collisions. Nevertheless, the spin polarized charged QGP serves as a paramagnet, which can efficiently polarize the strange quarks. This is realized through the splitting in the damping rates for quarks with opposite spin component along the magnetic field. Interestingly, this case leads to a splitting $\D\G\sim \frac{g^4T\m}{p}$, which is parametrically larger than the high density case with $\D\G\sim\frac{g^2T\quv^2}{\bmu p}\sim \frac{g^6T\m}{p}$.

Finally let us comment on two simplifications made in our analysis. The first one is the LLL approximation. For realistic magnetic field produced in heavy ion collisions, higher Landau level (hLL) contribution might not be negligible. Indeed, interesting phenomenological consequences have been discussed from inter-Landau level transitions \cite{Wang:2020dsr,Wang:2021ebh,Wang:2021eud}. How do they affect our results? Cutting the diagram in Fig.~\ref{fm1}, we easily see that the self-energy of the probe quark comes from scattering with either quarks in the LLL states or gluons. The dominance of either of the them correspond to the low and high density limits respectively. For the anti-symmetric part of self-energy leading to the splitting, we need at least one scattering with LLL states. It is crucial that the LLL states are spin polarized. The splitting in damping rates for opposite spin states seem to imply that it is the spin of the LLL states that are transferred to the counterpart of the probe fermions\footnote{Note again the spin is entirely canceled by the orbital angular momentum in the LLL approximation.}. If this were true, we would expect no qualitative change from hLL contributions, because hLL states are two-fold degenerate with opposite spins and their contributions would cancel in the splitting.

The second simplification is the neglection of quark mass, which is not necessarily small for strange quark as compared to temperature. One may ask when does our picture break down. For massive probe quark, the right and left handed components can convert to each other on time scale of axial charge relaxation. The time scale has been estimated as $\t_m\sim\frac{T}{m^2g^2}$ \cite{Hou:2017szz}. If the relaxation time scale is much longer than the splitting time scale $\t_{split}\sim \frac{1}{\D\G}$, the polarization mechanism we propose is still effective. The condition corresponds to $\frac{T}{m^2g^2}\gg \frac{p}{g^6T\m}$ in the high density limit and $\frac{T}{m^2g^2}\gg\frac{p}{g^4T\m}$ in the low density limit. If the relaxation time is much shorter than the splitting time scale instead, the polarization will be washed out.

\section{Conclusion and outlook}\label{co}

We have considered self-energies of photon/gluon in charged magnetized medium in strong magnetic field limit. Finite charge density of the medium induces an anti-symmetric component in the self-energies. We have found the anti-symmetric component leads to splitting of damping rates for probe chiral fermion/quark with opposite spin component along the magnetic field. We have analyzed the high and low density limits, finding medium behaves like a paramagnet in both cases. Applying the results to heavy ion collisions, we propose the QGP consisting of light quarks can be analogous to a paramagnet due to the interplay of finite magnetic field and baryon density. After decay of initial magnetic field, the paramagnet can continue to polarize the strange quarks produced at late stage of heavy ion collisions. This provides a mechanism to effectively extend life time of the magnetic field other than the electric conductivity.

Several extensions of this work can be considered: we have considered the strong magnetic field limit with the LLL approximation. We have argued the mechanism for polarization dynamics remains qualitatively the same based on the assumption that only the spin of the Landau level states matters for the polarization dynamics. It is desirable to confirm the picture in the weak field limit, where the spin and orbital angular momentum do not cancel each other. It is more interesting to consider the scenario with vorticity. Unlike the magnetized medium, the vortical medium carries mostly orbital angular momentum. It is known that an anti-symmetric component exists in self-energy for gluon in neutral vortical QGP \cite{Hou:2020mqp}. It is expected to lead to splitting of damping rates for different spin states, which leads to polarization mechanism through spin-orbit coupling. We leave these for future studies.

\section*{Acknowledgments}
We are grateful to Koichi Hattori, Aihong Tang and Pengfei Zhuang for helpful discussions. This work is in part supported by NSFC under Grant Nos 12075328 and 11735007.

\appendix


\section{Evaluation of the probe fermion self-energy}\label{q-integrations}
In this appendix, we evaluate the probe fermion self-energy which is necessary to determine the damping rate in the main text. Let's start from the anti-symmetric contribution of \eqref{srat}
\ba\label{rar}
\S_{ar}^{A}(P)&=&-\frac{ie^{2}T}{(2\p)^{3}\tm}\int \frac{d^{3}q}{q^2}\frac{f_{1}}{\bold{p}\cdot\bold{q}+i\e p_{0}}\no
&=&-\frac{ie^{2}T\g^{5}\g^{3}}{(2\p)^{3}\tm}\int \frac{d^{3}q}{q^2}\frac{p_{1}q_{1}+p_{2}q_{2}}{\bold{p}\cdot\bold{q}+i\e p_{0}}+\frac{ie^{2}T p_{3}\g^{5}}{(2\p)^{3}\tm}\int \frac{d^{3}q}{q^2}\frac{q_{1}\g^{1}+q_{2}\g^{2}}{\bold{p}\cdot\bold{q}+i\e p_{0}}+\frac{ie^{2}T\g^{5}\g^{3}}{(2\p)^{3}\tm}\int \frac{d^{3}q\ }{\bold{p}\cdot\bold{q}+i\e p_{0}}\no
&=&\frac{ie^{2}T\g^{5}\g^{3}}{(2\p)^{3}\tm}\int q_{\perp}dq_{\perp}d\ph\int\frac{dq_{3}}{p_{\perp} q_{\perp} cos{\ph}+p_{3}q_{3}+i\e p_{0}}\no
&-&\frac{ie^{2}T\g^{5}\g^{3}}{(2\p)^{3}\tm}p_{\perp}\int q_{\perp}^{2}cos{\ph}\ dq_{\perp}d\ph \int\frac{dq_{3}}{(q_{\perp}^{2}+q_{3}^{2})(p_{\perp} q_{\perp} cos{\ph}+p_{3}q_{3}+i\e p_{0})}
\no
&+&\frac{ie^{2}T p_{3}\g^{5}}{(2\p)^{3}\tm}\int q_{\perp}(\bold{q}_{\perp}\cdot\boldsymbol{\g_{\perp}}) \ dq_{\perp}d\ph \int\frac{dq_{3}}{(q_{\perp}^{2}+q_{3}^{2})(p_{\perp} q_{\perp} cos{\ph}+p_{3}q_{3}+i\e p_{0})},
\ea
with
\ba
\bold{q}_{\perp}\cdot\boldsymbol{\g_{\perp}}=\frac{q_{\perp}}{p_{\perp}}\((p_{1} cos\ph -p_{2}sin\ph )\g^{1}+(p_{1}sin\ph +p_{2}cos\ph )\g^{2}\).
\ea
$\ph$ is the angle between $\bold{p}_{\perp}$ and $\bold{q}_{\perp}$. We have used the cylindrical coordinates to calculate this integral.

Next, we will use the residue theorem to calculate the above integral. The sign of $p_{0}$ and $p_{3}$ will affect the integral result. Therefore, we consider the following two cases which are related to our study: one case is $p_{0}>0$ and $p_{3}>0$, the other case is $p_{0}>0$ and $p_{3}<0$.

As for the first case $(p_{0}>0\ and\ p_{3}>0)$, the integral results of $q_{3}$ are
\ba
\int\frac{dq_{3}}{p_{\perp} q_{\perp} cos{\ph}+p_{3}q_{3}+i\e p_{0}}=-\frac{i\pi}{p_{3}},\no
\int\frac{dq_{3}}{(q_{\perp}^{2}+q_{3}^{2})(p_{\perp} q_{\perp} cos{\ph}+p_{3}q_{3}+i\e p_{0})}=\frac{\p}{ q_{\perp}^{2}(ip_{3}+p_{\perp}cos\ph)},
\ea
Then, \eqref{rar} becomes
\ba
\S_{ar}^{A}(P)&=&\frac{ie^{2}T p_{3}q_{UV}}{8\p^{2}\tm p_{\perp}}\g^{5}\int d\ph \frac{(p_{1}cos\ph -p_{2}sin\ph )\g^{1}+(p_{1}sin\ph +p_{2}cos\ph )\g^{2}}{ip_{3}+p_{\perp}cos\ph}\no
&-&\frac{ie^{2}T p_{\perp} q_{UV}}{8\p^{2}\tm}\g^{5}\g^{3}\int\frac{ d\ph \ cos{\ph}}{ip_{3}+p_{\perp}cos\ph}+\frac{e^{2}T q_{UV}^{2}}{8\p\tm p_{3}}\g^{5}\g^{3}.
\ea
Let $z=e^{i\ph}$, $cos\ph=\frac{z^2+1}{2z}$ and $sin\ph=\frac{z^2-1}{2z}$, we can obtain
\ba\label{asf}
\S_{ar}^{A}(P)&=&\frac{e^{2}T p_{3} q_{UV} }{8\p^{2}\tm p_{\perp}}\g^{5}\oint_{\lvert z \rvert=1} \frac{dz}{z}\(\frac{(z^{2}+1)(p_{1}\g^{1}+p_{2}\g^{2})}{p_{\perp}(z^2+1)+2iz p_{3}}+\frac{(z^{2}-1)(p_{1}\g^{2}-p_{2}\g^{1})}{i(p_{\perp}(z^2+1)+2iz p_{3})}\)\no
&-&\frac{e^{2}T p_{\perp} q_{UV}}{8\p^{2}\tm}\g^{5}\g^{3}\oint_{\lvert z \rvert=1}\frac{dz}{z}\frac{z^{2}+1}{p_{\perp}(z^2+1)+2iz p_{3}}+\frac{e^{2}T q_{UV}^{2}}{8\pi\tm p_{3}}\g^{5}\g^{3},
\ea
The consequences of $\oint dz$ are
\ba\label{phi}
\oint_{\lvert z \rvert=1} \frac{dz}{z}\frac{(z^{2}+1)}{p_{\perp}(z^2+1)+2iz p_{3}}=\frac{2\pi i}{p_{\perp}}\(1-\frac{p_3}{p}\),\no
\oint_{\lvert z \rvert=1} \frac{dz}{z}\frac{(z^{2}-1)}{p_{\perp}(z^2+1)+2iz p_{3}}=0.
\ea
Substituting \eqref{phi} into \eqref{asf} , we can get the final result
\ba\label{self1}
\S_{ar}^{A}(P)=-\frac{ie^{2}T q_{UV}}{4\p\tm}\(1-\frac{p_{3}}{p}\)\(\g^{5}\g^{3}-\frac{p_{3}}{p_{\perp}^2}\g^{5}\(p_{1}\g^{1}+p_{2}\g^{2}\)\)+\frac{e^{2}T q_{UV}^{2}}{8\p\tm  p_{3}}\g^{5}\g^{3}.
\ea
We can use a similar method to calculate the second case $(p_{0}>0\ and \ p_{3}<0)$. The integral results of $q_{3}$ are
\ba
\int\frac{dq_{3}}{p_{\perp} q_{\perp} cos{\ph}+p_{3}q_{3}+i\e p_{0}}=\frac{i\pi}{p_{3}},\no
\int\frac{dq_{3}}{(q_{\perp}^{2}+q_{3}^{2})(p_{\perp} q_{\perp} cos{\ph}+p_{3}q_{3}+i\e p_{0})}=\frac{\p}{ q_{\perp}^{2}(p_{\perp}cos\ph-ip_{3}}).
\ea
Then, \eqref{rar} becomes
\ba
\S_{ar}^{A}(P)&=&\frac{ie^{2}T p_{3} q_{UV} }{8\p^{2}\tm p_{\perp}}\g^{5}\int d\ph \frac{(p_{1}cos\ph -p_{2}sin\ph )\g^{1}+(p_{1}sin\ph +p_{2}cos\ph )\g^{2}}{p_{\perp}cos\ph-ip_{3}}\no
&-&\frac{ie^{2}T p_{\perp} q_{UV} }{8\p^{2}\tm}\g^{5}\g^{3}\int\frac{ d\ph \ cos{\ph}}{p_{\perp}cos\ph-ip_{3}}-\frac{e^{2}T q_{UV}^{2}}{8\pi\tm p_{3}}\g^{5}\g^{3}.
\ea
We substitute $z=e^{i\ph}$, $cos\ph=\frac{z^2+1}{2z}$ and $sin\ph=\frac{z^2-1}{2z}$ into the above equation.
\ba
\S_{ar}^{A}(P)&=&\frac{e^{2}T p_{3} q_{UV} }{8\p^{2}\tm p_{\perp}}\g^{5}\oint_{\lvert z \rvert=1} \frac{dz}{z}\(\frac{(z^{2}+1)(p_{1}\g^{1}+p_{2}\g^{2})}{p_{\perp}(z^2+1)-2iz p_{3}}+\frac{(z^{2}-1)(p_{1}\g^{2}-p_{2}\g^{1})}{i(p_{\perp}(z^2+1)-2iz p_{3})}\)\no
&-&\frac{e^{2}T p_{\perp} q_{UV}}{8\p^{2}\tm}\g^{5}\g^{3}\oint_{\lvert z \rvert=1}\frac{dz}{z}\frac{z^{2}+1}{p_{\perp}(z^2+1)-2iz p_{3}}-\frac{e^{2}T q_{UV}^{2}}{8\p\tm p_{3}}\g^{5}\g^{3}.
\ea
The consequences of $\oint dz$ are
\ba
\oint_{\lvert z \rvert=1} \frac{dz}{z}\frac{z^{2}+1}{p_{\perp}(z^2+1)-2iz p_{3}}=\frac{2\pi i}{p_{\perp}} \(1+\frac{p_{3}}{p}\),\no
\oint_{\lvert z \rvert=1} \frac{dz}{z}\frac{z^{2}-1}{(p_{\perp}(z^2+1)-2iz p_{3})}=0.
\ea
Eventually, we get
\ba\label{self11}
\S_{ar}^{A}(P)=-\frac{ie^{2}T q_{UV}}{4\p\tm}\(1+\frac{p_3}{p}\)\(\g^{5}\g^{3}-\frac{p_{3}}{p_{\perp}^2}\g^{5}\(p_{1}\g^{1}+p_{2}\g^{2}\)\)-\frac{e^{2}T q_{UV}^{2}}{8\p\tm  p_{3}}\g^{5}\g^{3}.
\ea

Combining \eqref{self1} and \eqref{self11}, we can back to the result of \eqref{sarmm}.

Let's turn to the symmetric contribution. We start from the following expression
\ba\label{syself}
\S_{ar}^{S}(P)&=&\frac{e^{2}}{(2\p)^{3}}\int \frac{d^{4}Q\ \sgn(q_{0})}{(P-Q)^2+i\e(p_{0}-q_{0})}\(\frac{1}{2}+f_{\g}(q_{0})\)\(\frac{\d(q_{0}^{2}-x_{1}^{2})}{q_{0}^{2}-x_{2}^{2}}+\frac{\d(q_{0}^{2}-x_{2}^{2})}{q_{0}^{2}-x_{1}^{2}}\)\no
&\times&\g^{\m}(\slashed{P}-\slashed{Q})\g^{\n}S_{\m\n}(Q),
\ea
We first deal with $\g^{\m}(\slashed{P}-\slashed{Q})\g^{\n}S_{\m\n}(Q)$. The following result is obtained by considering only $\sim q_{0}^{0}$
\ba
\g^{\m}(\slashed{P}-\slashed{Q})\g^{\n}S_{\m\n}(Q)=-2iq_{3}^{2}\(p_{0}\g^{0}+\bold{q}\cdot\boldsymbol{\g}-\frac{(\bold{q}\cdot\boldsymbol{\g})(\bold{p}\cdot\bold{q})}{q^{2}}\).
\ea
Then, \eqref{syself} can be written as
\ba\label{ssra0}
\S_{ar}^{S}(P)=\frac{ie^{2}T}{(2\p)^{3}\tm^{2}}\int \frac{q_{3}^{2}\(p_{0}\g^{0}+\bold{q}\cdot\boldsymbol{\g}-\frac{(\bold{q}\cdot\boldsymbol{\g})(\bold{p}\cdot\bold{q})}{q^{2}}\)}{\(\bold{p}\cdot\bold{q}+i\e p_{0}\)}d^{3}q\int \frac{d q_{0}}{q_{0}}\d\(q_{0}^{2}-\frac{q_{3}^{2}q^{2}}{\tm^{2}}\)\sgn(q_{0}).
\ea
We proceed with the integral of $q_{0}$.
\ba\label{sq0}
\int \frac{d q_{0}}{q_{0}}\d\(q_{0}^{2}-\frac{q_{3}^{2}q^{2}}{\tm^{2}}\)\sgn(q_{0})=\frac{\tm^{2}}{q_{3}^{2}q^{2}}.
\ea
By using \eqref{sq0}, we can simplify \eqref{ssra0} and obtain
\ba
\S_{ar}^{S}(P)=\frac{ie^{2}T}{(2\p)^{3}}\int \frac{d^{3}q}{q^{2}\(\bold{p}\cdot\bold{q}+i\e p_{0}\)}\(p_{0}\g^{0}+\bold{q}\cdot\boldsymbol{\g}-\frac{(\bold{q}\cdot\boldsymbol{\g})(\bold{p}\cdot\bold{q})}{q^{2}}\).
\ea
In cylindrical coordinates, the above equation can be rewritten as
\ba\label{ssff}
\S_{ar}^{S}(P)
&=&\frac{ie^{2}T\g^{0}p0}{(2\p)^{3}}\int q_{\perp}dq_{\perp}d\ph\int\frac{dq_{3}}{(q_{\perp}^{2}+q_{3}^{2})(p_{\perp} q_{\perp} cos{\ph}+p_{3}q_{3}+i\e p_{0})}\no
&+&\frac{ie^{2}T}{(2\p)^{3}}\int q_{\perp}(\bold{q}_{\perp}\cdot\boldsymbol{\g_{\perp}})\ dq_{\perp}d\ph \int\frac{dq_{3}}{(q_{\perp}^{2}+q_{3}^{2})(p_{\perp} q_{\perp} cos{\ph}+p_{3}q_{3}+i\e p_{0})}\no
&+&\frac{ie^{2}T\g^{3}}{(2\p)^{3}}\int q_{\perp}\ dq_{\perp}d\ph \int\frac{q_{3} dq_{3}}{(q_{\perp}^{2}+q_{3}^{2})(p_{\perp} q_{\perp} cos{\ph}+p_{3}q_{3}+i\e p_{0})}\no
&-&\frac{ie^{2}T}{(2\p)^{3}}\int q_{\perp}(\bold{q}_{\perp}\cdot\boldsymbol{\g_{\perp}})(\bold{q}_{\perp}\cdot\bold{p}_{\perp})\ dq_{\perp}d\ph \int\frac{dq_{3}}{(q_{\perp}^{2}+q_{3}^{2})^{2}(p_{\perp} q_{\perp} cos{\ph}+p_{3}q_{3}+i\e p_{0})}\no
&-&\frac{ie^{2}T p_{3}}{(2\p)^{3}}\int q_{\perp}(\bold{q}_{\perp}\cdot\boldsymbol{\g_{\perp}})\ dq_{\perp}d\ph \int\frac{q_{3}dq_{3}}{(q_{\perp}^{2}+q_{3}^{2})^{2}(p_{\perp} q_{\perp} cos{\ph}+p_{3}q_{3}+i\e p_{0})}\no
&-&\frac{ie^{2}T \g_{3}}{(2\p)^{3}}\int q_{\perp}(\bold{q}_{\perp}\cdot\bold{p_{\perp}})\ dq_{\perp}d\ph \int\frac{q_{3}dq_{3}}{(q_{\perp}^{2}+q_{3}^{2})^{2}(p_{\perp} q_{\perp} cos{\ph}+p_{3}q_{3}+i\e p_{0})}\no
&-&\frac{ie^{2}T \g_{3}p_{3}}{(2\p)^{3}}\int q_{\perp}\ dq_{\perp}d\ph \int\frac{q_{3}^{2} dq_{3}}{(q_{\perp}^{2}+q_{3}^{2})^{2}(p_{\perp} q_{\perp} cos{\ph}+p_{3}q_{3}+i\e p_{0})}.
\ea
We still consider two cases. As for the case of $p_{0}>0$ and $p_{3}>0$, the integral results of $q_{3}$ are
\ba\label{dq3}
\int\frac{q_{3} dq_{3}}{(q_{\perp}^{2}+q_{3}^{2})(p_{\perp} q_{\perp} cos{\ph}+p_{3}q_{3}+i\e p_{0})}=\frac{i\p}{ q_{\perp}(ip_{3}+p_{\perp}cos\ph)},\no
\int\frac{dq_{3}}{(q_{\perp}^{2}+q_{3}^{2})^{2}(p_{\perp} q_{\perp} cos{\ph}+p_{3}q_{3}+i\e p_{0})}=\frac{\p(2ip_{3}+p_{\perp}cos\ph)}{2 q_{\perp}^{4}(ip_{3}+p_{\perp}cos\ph)^{2}},\no
\int\frac{q_{3}dq_{3}}{(q_{\perp}^{2}+q_{3}^{2})^{2}(p_{\perp} q_{\perp} cos{\ph}+p_{3}q_{3}+i\e p_{0})}=\frac{-p_{3}\p}{2 q_{\perp}^{3}(ip_{3}+p_{\perp}cos\ph)^{2}},\no
\int\frac{q_{3}^{2} dq_{3}}{(q_{\perp}^{2}+q_{3}^{2})^{2}(p_{\perp} q_{\perp} cos{\ph}+p_{3}q_{3}+i\e p_{0})}=\frac{\p p_{\perp}cos\ph}{2 q_{\perp}^{2}(ip_{3}+p_{\perp}cos\ph)^{2}},
\ea
We plug \eqref{dq3} into \eqref{ssff} and calculate the integral of $q_{\perp}$ to get the following result.
\ba\label{ssfff}
\S_{ar}^{S}(P)&=&\frac{ie^{2}T \g^{0}p_{0}}{8\p^{2}}\ln\frac{\quv}{\qir}\int \frac{d\ph}{ip_{3}+p_{\perp}cos\ph}-\frac{e^{2}T \g^{3}q_{UV}}{8\p^{2}}\int\frac{d\ph}{ip_{3}+p_{\perp}cos\ph}\no
&+&\frac{ie^{2}T q_{UV} }{8\p^{2}p_{\perp}}\int\frac{d\ph}{ip_{3}+p_{\perp}cos\ph}\(cos\ph(\g^{1}p_{1}+\g^{2}p_{2})+sin\phi(\g^{2}p_{1}-\g^{1}p_{2})\)\no
&-&
\frac{ie^{2}T}{16\p^{2}}\ln\frac{\quv}{\qir}\int \frac{(2ip_{3}+p_{\perp}cos\ph)d\ph}{(ip_{3}+p_{\perp}cos\ph)^{2}}\((cos\ph)^{2}(\g^{1}p_{1}+\g^{2}p_{2})+sin\ph cos\ph(\g^{2}p_{1}-\g^{1}p_{2})\)\no
&+&\frac{ie^{2}T p_{3}^{2}}{16\p^{2}p_{\perp}}\ln\frac{\quv}{\qir}\int \frac{d\ph}{(ip_{3}+p_{\perp}cos\ph)^{2}}\(cos\ph(\g^{1}p_{1}+\g^{2}p_{2})+sin\ph(\g^{2}p_{1}-\g^{1}p_{2})\).
\ea
Then we can calculate the integral of $\ph$ and write as
\ba\label{sphi}
\int \frac{d\ph}{ip_{3}+p_{\perp}cos\ph}=\frac{-2i\p}{p},\no
\int\frac{\(cos\ph(\g^{1}p_{1}+\g^{2}p_{2})+sin\phi(\g^{2}p_{1}-\g^{1}p_{2})\)d\ph}{ip_{3}+p_{\perp}cos\ph}=\frac{2\p}{p_{\perp}}\(1-\frac{p_{3}}{p}\)(p^{1}\g_{1}+p_{2}\g^{2}),\no
\int \frac{d\ph}{(ip_{3}+p_{\perp}cos\ph)^{2}}\(cos\ph(\g^{1}p_{1}+\g^{2}p_{2})+sin\ph(\g^{2}p_{1}-\g^{1}p_{2})\)=-2i\p\frac{p_{\perp}}{p^{3}}(\g^{1}p_{1}+\g^{2}p_{2}),\no
\int \frac{(2ip_{3}+p_{\perp}cos\ph)d\ph}{(ip_{3}+p_{\perp}cos\ph)^{2}}\((cos\ph)^{2}(\g^{1}p_{1}+\g^{2}p_{2})+sin\ph cos\ph(\g^{2}p_{1}-\g^{1}p_{2})\)=-2i\p\frac{ p_{3}^{2}}{p^{3}}(\g^{1}p_{1}+\g^{2}p_{2}),\no
\ea
Substituting \eqref{sphi} into \eqref{ssfff}, we can obtain
\ba\label{aself1}
\S_{ar}^{S}(P)&=&\frac{e^{2}T \g^{0}p_{0}}{4\p p}\ln\frac{\quv}{\qir}+\frac{ie^{2}T \g^{3}q_{UV}}{4\p^{2}p}+\frac{ie^{2}T q_{UV}}{4\p p_{\perp}^{2}}\(1-\frac{p_{3}}{p}\)(\g^{1}p_{1}+\g^{2}p_{2})
\ea

As for the another case of $p_{0}>0$ and $p_{3}<0$, the integral results of $q_{3}$ change into
\ba\label{ssqq3}
\int\frac{q_{3} dq_{3}}{(q_{\perp}^{2}+q_{3}^{2})(p_{\perp} q_{\perp} cos{\ph}+p_{3}q_{3}+i\e p_{0})}=\frac{-i\p}{ q_{\perp}(p_{\perp}cos\ph-ip_{3})},\no
\int\frac{dq_{3}}{(q_{\perp}^{2}+q_{3}^{2})^{2}(p_{\perp} q_{\perp} cos{\ph}+p_{3}q_{3}+i\e p_{0})}=\frac{\p(p_{\perp}cos\ph-2ip_{3})}{2 q_{\perp}^{4}(p_{\perp}cos\ph-ip_{3})^{2}},\no
\int\frac{q_{3}dq_{3}}{(q_{\perp}^{2}+q_{3}^{2})^{2}(p_{\perp} q_{\perp} cos{\ph}+p_{3}q_{3}+i\e p_{0})}=\frac{-p_{3}\p}{2 q_{\perp}^{3}(p_{\perp}cos\ph-ip_{3})^{2}},\no
\int\frac{q_{3}^{2} dq_{3}}{(q_{\perp}^{2}+q_{3}^{2})^{2}(p_{\perp} q_{\perp} cos{\ph}+p_{3}q_{3}+i\e p_{0})}=\frac{\p p_{\perp}cos\ph}{2 q_{\perp}^{2}(p_{\perp}cos\ph-ip_{3})^{2}}.
\ea
After using the result of \eqref{ssqq3}, \eqref{ssff} becomes
\ba
\S_{ar}^{S}(P)&=&\frac{ie^{2}T \g^{0}p_{0}}{8\p^{2}}\ln\frac{\quv}{\qir}\int \frac{d\ph}{p_{\perp}cos\ph-ip_{3}}+\frac{e^{2}T \g^{3}q_{UV}}{8\p^{2}}\int\frac{d\ph}{p_{\perp}cos\ph-ip_{3}}\no
&+&\frac{ie^{2}T q_{UV} }{8\p^{2}p_{\perp}}\int\frac{d\ph}{p_{\perp}cos\ph-ip_{3}}\(cos\ph(\g^{1}p_{1}+\g^{2}p_{2})+sin\phi(\g^{2}p_{1}-\g^{1}p_{2})\)\no
&-&
\frac{ie^{2}T}{16\p^{2}}\ln\frac{\quv}{\qir}\int \frac{(p_{\perp}cos\ph-2ip_{3})d\ph}{(p_{\perp}cos\ph-ip_{3})^{2}}\((cos\ph)^{2}(\g^{1}p_{1}+\g^{2}p_{2})+sin\ph cos\ph(\g^{2}p_{1}-\g^{1}p_{2})\)\no
&+&\frac{ie^{2}T p_{3}^{2}}{16\p^{2}p_{\perp}}\ln\frac{\quv}{\qir}\int \frac{d\ph}{(p_{\perp}cos\ph-ip_{3})^{2}}\(cos\ph(\g^{1}p_{1}+\g^{2}p_{2})+sin\ph(\g^{2}p_{1}-\g^{1}p_{2})\).
\ea
We take advantage of the same method as before to integrate $\ph$ to get the following result.
\ba\label{aself11}
\int \frac{d\ph}{p_{\perp}cos\ph-ip_{3}}=\frac{-2i\p}{p}\no
\int\frac{\(cos\ph(\g^{1}p_{1}+\g^{2}p_{2})+sin\phi(\g^{2}p_{1}-\g^{1}p_{2})\)d\ph}{p_{\perp}cos\ph-ip_{3}}=\frac{2\p}{p_{\perp}}\(1+\frac{p_{3}}{p}\)(p^{1}\g_{1}+p_{2}\g^{2})\no
\int \frac{d\ph}{(p_{\perp}cos\ph-ip_{3})^{2}}\(cos\ph(\g^{1}p_{1}+\g^{2}p_{2})+sin\ph(\g^{2}p_{1}-\g^{1}p_{2})\)=-2i\p\frac{p_{\perp}}{p^{3}}(\g^{1}p_{1}+\g^{2}p_{2})\no
\int \frac{(p_{\perp}cos\ph-2ip_{3})d\ph}{(p_{\perp}cos\ph-ip_{3})^{2}}\((cos\ph)^{2}(\g^{1}p_{1}+\g^{2}p_{2})+sin\ph cos\ph(\g^{2}p_{1}-\g^{1}p_{2})\)=-2i\p\frac{ p_{3}^{2}}{p^{3}}(\g^{1}p_{1}+\g^{2}p_{2})\no
\ea
In the end, we can obtain
\ba\label{aself11d}
\S_{ar}^{S}(P)&=&\frac{e^{2}T \g^{0}p_{0}}{4\p p}\ln\frac{\quv}{\qir}-\frac{ie^{2}T \g^{3}q_{UV}}{4\p^{2}p}+\frac{ie^{2}T q_{UV}}{4\p p_{\perp}^{2}}\(1+\frac{p_{3}}{p}\)(\g^{1}p_{1}+\g^{2}p_{2})
\ea

Combining \eqref{aself1} and \eqref{aself11d}, we can obtain the result of \eqref{asarmm}. Similarly, for the case where $p_{0}<0$, we do not elaborate further.

\bibliographystyle{unsrt}
\bibliography{QGP_magnet}

\end{document}